\documentclass[review]{elsarticle}

\usepackage{amsmath}
\usepackage[T1]{fontenc}
\usepackage{changepage}
\usepackage{graphicx}
\usepackage[hidelinks]{hyperref}
\usepackage{lmodern}
\usepackage{multirow}
\usepackage{subcaption}
\usepackage[disable]{todonotes}

\usepackage{listings}
\graphicspath{{./figures/}}
\DeclareGraphicsExtensions{.pdf}
\journal{Computer Languages, Systems and Structures}
\usepackage[absolute]{textpos}
\makeatletter
\def\ps@pprintTitle{
\def\@oddfoot{}
}
\makeatother

\begin{document}

\begin{textblock}{15}(0.5,14.5)
{
\vspace{-2pt}\noindent\hrulefill

\noindent\fontsize{8pt}{8pt}\selectfont\copyright\ 2016. This manuscript version is made available under the CC-BY-NC-ND 4.0 license: \url{http://creativecommons.org/licenses/by-nc-nd/4.0/}. \hspace{5pt} This is the accepted version of: M. Sul\'ir, M. Nos\'a\v{l}, J. Porub\"an. Recording concerns in source code using annotations. Computer Languages, Systems and Structures (COMLAN), Vol. 46, 2016, pp. 44--65, Elsevier. \url{http://doi.org/10.1016/j.cl.2016.07.003}

}
\end{textblock}

\begin{frontmatter}

\title{Recording Concerns in Source Code Using Annotations}

\author{Mat\'u\v{s} Sul\'ir}
\ead{matus.sulir@tuke.sk}
\author{Milan Nos\'a\v{l}}
\ead{milan.nosal@gmail.com}
\author{Jaroslav Porub\"an}
\ead{jaroslav.poruban@tuke.sk}
\address{Department of Computers and Informatics\\
Faculty of Electrical Engineering and Informatics\\
Technical University of Ko\v{s}ice\\
Letn\'a 9, 042 00 Ko\v{s}ice, Slovakia}

\begin{abstract}

A concern can be characterized as a developer's intent behind a piece of code, often not explicitly captured in it. We discuss a technique of recording concerns using source code annotations (concern annotations). Using two studies and two controlled experiments, we seek to answer the following 3 research questions: 1) Do programmers' mental models overlap? 2) How do developers use shared concern annotations when they are available? 3) Does using annotations created by others improve program comprehension and maintenance correctness, time and confidence? The first study shows that developers' mental models, recorded using concern annotations, overlap and thus can be shared. The second study shows that shared concern annotations can be used during program comprehension for the following purposes: hypotheses confirmation, feature location, obtaining new knowledge, finding relationships and maintenance notes. The first controlled experiment with students showed that the presence of annotations significantly reduced program comprehension and maintenance time by 34\%. The second controlled experiment was a differentiated replication of the first one, focused on industrial developers. It showed a 33\% significant improvement in correctness. We conclude that concern annotations are a viable way to share developers' thoughts.
\end{abstract}


\end{frontmatter}


\section{Introduction}

Programmers developing a program continuously create a mental model, which is a representation of the program in their mind. They try to express the mental model in the programming language constructions. However, some parts of the mental model are not explicitly expressed in the source code. They are either implicitly indicated in complicated implementation details, or lost.

\subsection{Motivation}

\begin{figure}
\centering
\includegraphics[width=0.85\hsize]{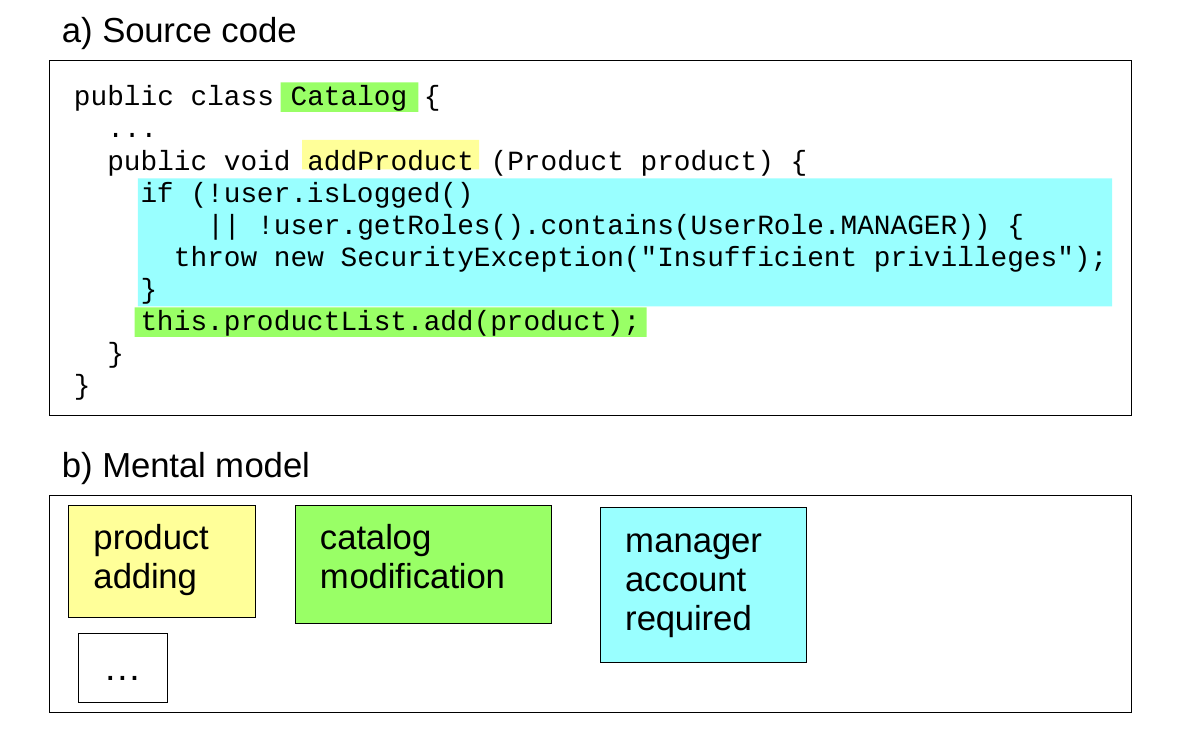}
\caption{An example of unannotated source code and the corresponding mental model} \label{fig:model}
\end{figure}

Suppose a developer encounters the Java code excerpt shown in Figure~\ref{fig:model} a). While reading it, the following ideas may gradually appear in his or her mind:
\begin{itemize}
\item The method \texttt{addProduct} adds some product somewhere.
\item As it is in the \texttt{Catalog} class, it adds the product to the catalog.
\item An addition to a catalog changes its state -- the catalog is modified.
\item There is a condition checking whether a user is logged.
\item To successfully perform this operation, a logged user must be a manager; otherwise an exception is thrown.
\end{itemize}

While this process may seem natural, it has the following shortcomings:
\begin{itemize}
\item The fact that the code modifies a catalog is formed only after reading two parts of the code far apart.
\item The information that a ``manager account'' is necessary is formed after reading a rather lengthy piece of code.
\item There is one important piece of information not identified at all: The added product actually appears in the catalog only after the administrator approves it. This information is not apparent from this excerpt, and may be the result of cooperation of many remotely related classes.
\end{itemize}

\begin{figure}
\centering
\includegraphics[width=0.85\hsize]{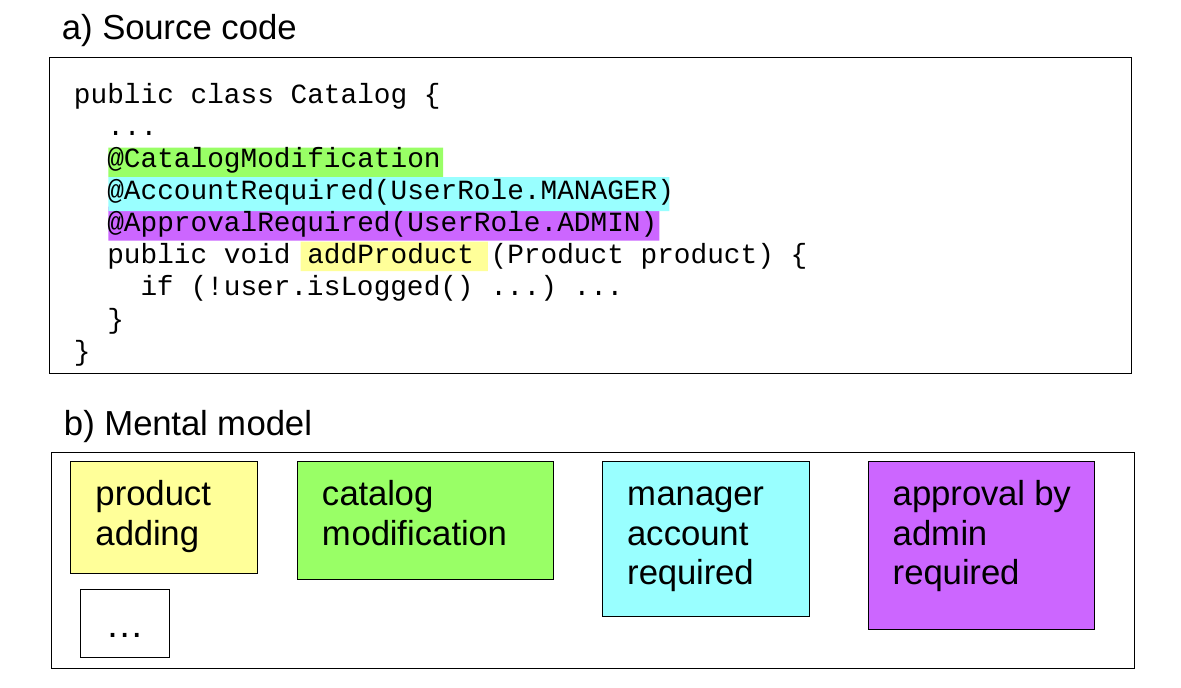}
\caption{The annotated source code and the corresponding mental model} \label{fig:annotations}
\end{figure}

Now suppose the developer originally writing the code, or any other programmer maintaining it, annotated the source code with concern annotations (source code annotations are declarative marks used to decorate source code with metadata \cite{fedcsisAnnLan}), as displayed in Figure~\ref{fig:annotations}~a). The developer later reading it would rapidly obtain a mental model containing useful information about the code, without even looking at the method implementation, scrolling through the code to stitch the pieces of information, or even debug it to explain unexpected behavior. Furthermore, it would be possible to find all methods requiring approval in case of refactoring or business policy changes. As annotations are a part of the Java language, no additional tools are required: standard IDE (integrated development environment) features like Find Usages could be used.

\subsection{Aim}

The purpose of this paper is to investigate the viability of source code annotations as a medium to share parts of developers' mental models. Therefore, we form our \textbf{hypothesis} as follows: Annotations created by one group of developers are useful for comprehension and maintenance tasks performed by other developers.

Our first goal is to find out whether at least some of the concerns recognized by one programmer can be recognized by other developers. If so, then the mental model of one developer at least partially overlaps with the other persons' mental model. This is a necessary condition for concern annotation sharing to be useful. Then we examine the usability of concern annotations using an observational study and two controlled experiments.

We formulate the research questions as follows:

\begin{itemize}
\item \textbf{RQ1}: Do programmers' mental models, recorded using concern annotations, overlap?
\item \textbf{RQ2}: How do developers use shared annotations when they are available?
\item \textbf{RQ3}: Can annotations created by others improve program comprehension and maintenance correctness, time and confidence?
\end{itemize}

To answer each of the questions, we used an appropriate empirical research method \cite{Wohlin12experimentation}.


\section{Basic Concepts}
\label{sec:concernAnnotations}

Before describing the studies, we introduce the notion of concerns, concern annotations, and their relation to a mental model.

\subsection{Mental Model}

A mental model is ``an internal, working representation of the software under consideration'' \cite{Mayrhauser95program}. It is described by its static structures -- pieces of knowledge, and by dynamic behavior -- the processes forming these pieces. Now we will focus mainly on the static structures.

Since there have been multiple attempts to describe the structure of a mental model, each researcher has its own point of view. Mayrhauser and Vans distinguish these types of structures in their review \cite{Mayrhauser95program}: \textit{Chunks} are mappings from a label like ``sorting'' to the program text implementing it, or to a set of other chunks like ``comparing'' and ``swapping''. \textit{Plans} are similar, but they may include relationships across various pieces of the code, and the problem domain knowledge. \textit{Hypotheses} are conjectures about parts of the program, which the programmers try to prove or disprove.

Our view of a mental model unifies chunks, plans and hypotheses into a set of concerns. In the programmer's mind, a concern is represented by a short label, as shown in Figure~\ref{fig:model} b), and its corresponding source code elements.

\subsection{Concerns}

A \textit{concern} can be characterized as a developer's intent of a particular piece of code: What should this code accomplish? What am I trying to achieve with it? How would I tersely characterize it? Is there something special about it? Some concerns may be obvious by looking at the code itself (chiefly from the identifiers), but many concerns are hidden. Basically, a concern represents a software system requirement, or a design decision about the code \cite{greenfield}.

A concern can be observed in three dimensions \cite{vranicIntent}: \emph{stakeholders} (from whose perspective is the concern observed -- e.g., a programmer); \emph{level} (at what level of construct granularity is the concern expressed, e.g., low-level constructs such as a loop); and \emph{expression} (how is the concern expressed, starting with the mental model and ending with the source code).

Due to these and other reasons, concerns' coverage can overlap. For example, a single method can implement persistence -- contributing to the persistence concern; and in the same time it can contain code securing it against unauthorized access -- a part of the security concern implementation. Thus, one piece of code can belong to multiple concerns \cite{Poruban14leveraging}, in this case, both persistence and security.

\subsection{Concern Annotations}

Since a method can belong to multiple concerns, expressing concerns using only identifiers (naming conventions) is not sufficient: naming conventions do not scale well. See Listing~\ref{lst:conventions} for an example using the Persistence keyword in the identifier to express that the method implements persistence, and SecuredMethod with OnlyAdmin to specify that it contributes to the security concern by allowing access only to an administrator.

\begin{lstlisting}[caption={Expressing concerns using naming conventions}, label={lst:conventions}]
public addNewUserPersistenceSecuredMethodOnlyAdmin(User user) { ... }
\end{lstlisting}

It is possible for a class or method in Java to have more than one annotation, scaling well when used for expressing concerns. Therefore, source code annotations (attributes in C\# terminology) are a better alternative to identifiers for expressing concerns directly in the source code. Listing~\ref{lst:annotations} shows the example from Listing~\ref{lst:conventions}, rewritten using concern annotations. For the persistence concern, we used the \texttt{@Persistence} annotation; for the security concern, we used the \texttt{@SecuredMethod} annotation. To specify which roles are allowed to access the method, we used the \texttt{allowed} parameter of the \texttt{@SecuredMethod} annotation.

\begin{lstlisting}[caption={Concern annotations}, label={lst:annotations}]
@Persistence
@SecuredMethod(allowed={User.ADMIN})
public addNewUser(User user) { ... }
\end{lstlisting}

For each distinct concern, we recommend creating one Java \textit{annotation type}. For example, we can create an annotation type \texttt{Persistence} which tells us the code marked with it fulfills the task of persistent object storage. Subsequently, we can mark the methods such as \texttt{addNewUser()} or \texttt{deleteUser()}, and the class \texttt{Database} with it. We will call them \textit{annotation occurrences}.

\subsection{Concern Annotations Properties}
\label{sec:concernAnnotProp}

Compared to traditional source code comments, concern annotations are more formal. We can use standard IDE features like navigating to the declaration, usages searching, refactoring, and other on them. However, we want to note that concern annotations are not meant to substitute traditional comments, but rather to complement them. Furthermore, annotation type definitions may be also commented by natural language comments if needed.

Still, we expect annotations to have some advantages over comments when considered from the viewpoint of readability. For example, let us consider the thread safety concern and \texttt{javax.swing.JPanel}\footnote{ \url{http://docs.oracle.com/javase/8/docs/api/javax/swing/JPanel.html}} in listing~\ref{lst:commentJP}. Thanks to its brevity, the \texttt{@NotThreadSafe} annotation is easier to spot than its natural language counterpart in the JavaDoc comment (even though we omitted the rest of the comment). This is because the annotation compresses two natural language sentences into a single identifier -- \texttt{NotThreadSafe}. Of course, the comment provides the same information, however, due to its brevity, the annotation is easier to spot and comprehend. For a more detailed discussion on annotations and comments, we refer the reader to our previous work \cite{poruban2014reusable}. 

\begin{lstlisting}[caption={The JPanel JavaDoc comment and the annotation}, label={lst:commentJP}]
/**
 * ... comments ...
 * <strong>Warning:</strong> Swing is not thread safe. For more
 * information see <a
 * href="package-summary.html#threading">Swing's Threading
 * Policy</a>.
 * ... comments ...
 */
@NotThreadSafe
public class JPanel extends JComponent implements Accessible
{
\end{lstlisting}

Often a concern annotation is just a simple label without a structure -- like the mentioned \texttt{\makeatletter{}@\makeatother{}NotThreadSafe} annotation. However, in some cases, parameters may be convenient. To avoid natural language usage, enumeration type parameters are preferred to strings. See the \texttt{\makeatletter{}@\makeatother{}SecuredMethod} annotation in Listing~\ref{lst:annotations} for an example. As we will see in section~\ref{sec:overlappingProcedure}, developers adhere to this rule.

\subsection{Kinds of Concern Annotations}
\label{sec:kindsConcern}

\textit{Domain annotations} document concepts and features of the application (problem) domain. For example, all source code elements representing the feature of filtering a list of items can be marked with the annotation \texttt{\makeatletter{}@\makeatother{}Filtering}. Similarly, all code related to bibliographic citations could be annotated by \texttt{\makeatletter{}@\makeatother{}Citing}.

\textit{Design annotations} document design and implementation decisions like design patterns, e.g., \texttt{\makeatletter{}@\makeatother{}Observer}.

\textit{Maintenance annotations} are intended to replace the traditional TODO and related comments. An example is the \texttt{\makeatletter{}@\makeatother{}Unused} annotation for parts of code not used in the project at all.

\subsection{Limitations of Concern Annotations}

There are some limitations of annotations as a technique for expressing concerns.

The first one is their granularity. Annotations annotate identifiable program elements -- thus, they can be used to express concerns at the covering constructs level (methods, classes), partially low level constructs (variables, parameters), and partially conceptual constructs (subsystems represented by packages that can be annotated) as well. However, they cannot be used to associate every line of code with its concern.

Another limitation is the flat structure of annotations. Each concern is expressed only by an annotation name, and optionally also by its parameters. Although annotations support nesting (an annotation within an annotation), they do not support cycles (general tree structures). Right now, annotation type inheritance is not supported, which limits the options to structure concern annotations. Nevertheless, annotations can denote an ``is-a'' relationship between concerns, like ``Observer is-a DesignPattern'', if desired. This is possible thanks to meta-annotations\footnote{Meta-annotations are annotations that annotate annotation types. There are several built-in meta-annotations in Java, such as {\scriptsize\texttt{\makeatletter{}@\makeatother{}Target}}, which is used to annotate annotation types to indicate the kinds of program elements to which the annotation type is applicable. If for any annotation type its {\scriptsize\texttt{\makeatletter{}@\makeatother{}Target}} annotation includes the {\scriptsize\texttt{ANNOTATION\_TYPE}} element kind, it means that the given annotation type is applicable to other element types -- meaning it is a custom meta-annotation type.}: We could annotate the annotation type definition of {\small\texttt{Observer}} with a meta-annotation {\small\texttt{\makeatletter{}@\makeatother{}DesignPattern}}.


\section{Mental Model Overlapping}
\label{sec:modelOverlapping}

In the first study, we examine whether programmers' mental models recorded using concern annotations overlap.

Note that by ``developers'' (or its synonym, ``programmers''), we mean both the original creator of the program, and the persons reading it or trying to understand some parts in order to maintain it. Nowadays, software development is a highly collaborative process and the distinction between a standard programmer and a maintenance programmer is practically nonexistent.

\subsection{Method}

We asked multiple programmers to independently annotate the source code of an existing program with concern annotations. Then we measured to what extent their concern annotations overlap.

\subsubsection{Materials}

For this and subsequent studies, we used EasyNotes\footnote{\url{http://github.com/MilanNosal/easy-notes}} -- a desktop application for bibliographic note-taking. It is a small-scale Java project consisting of around 2500 lines of code located in 33 classes. Except for scarce source code comments, it has no documentation available.

\subsubsection{Participants}

This study had 7 participants:
\begin{description}
\item[A] a researcher and lecturer with a PhD in Computer Science,
\item[B] an industrial Java programmer,
\item[C] a postdoctoral researcher and Java programmer,
\item[D] an associate professor with extensive Java experience,
\item[E, F] first-year PhD students,
\item[G] the author of EasyNotes.
\end{description}

The author knew the program well, since he was its sole author. The other subjects had performed a small maintenance task on it in the past.

The activity was individual and the participants were not allowed to interact during the experiment.

\subsubsection{Procedure}
\label{sec:overlappingProcedure}

First, the participants were given the original source code of EasyNotes without any annotations (commit {\small\texttt{a299e64}}). They had an arbitrary amount of time available to become familiar with the application both from an end-user and programmer perspective.

Next, they were asked to create a custom annotation type for each concern they recognized in the application and to mark the classes, member variables and methods with the annotations they just created whenever they considered it appropriate. Participants were not given any special instructions on how to recognize the concerns, nor how to design corresponding annotation types. We decided to give them this freedom so that we could get the best insight into how they would represent their mental model using concern annotations. Furthermore, the participants were asked to comment each annotation type to explain the concern it represents in natural language.

Participants were recommended to use either marker annotations (without parameters, only the name of the annotation is significant), or to use parameters that are not strings (strings would allow using natural language inside of parameters). This recommendation aimed to keep annotations as structured as possible -- to avoid parsing of arbitrary strings. In the end, most of the annotation types were just marker annotations. The exceptions included several enumeration parameters (with between 2 and 4 enumeration constants), boolean parameters, and several string parameters. All usages of string parameters were justified; all of them were used in maintenance annotations that required an arbitrary text to be associated with them. For instance, a TODO concern annotation with a task that needed to be done, a Question About Code concern annotation with a particular question, etc. Examples of such parameters can be found in the commit \texttt{f52872b}.

The participants were not limited in the time they needed for comprehending and annotating the code, and the time was not measured. However, all the subjects managed to finish the job approximately between 1 and 1.75 hours.

\subsubsection{Analysis}
\label{sec:analysisFirstExp}

Finally, we collected the modified projects and analyzed them semi-automati\-cally, looking for an overlap in annotation types and the use of the annotations on specific elements.

First, we identified concerns expressed by concern annotations. This was done by the second author of this paper, who manually inspected all concern annotations in each version of the EasyNotes project annotated during the experiment. The process consisted of a manual inspection of annotation types and their natural language comments. For each annotation type, we performed a further inspection of its annotations in the code to check whether the code annotated with concern annotations was consistent with our understanding of the concern. In ambiguous cases, we followed up by having a discussion with the participant -- the author of the concern in question. This was also the reason why we decided not to use automatic matching based on ontologies. This thorough process was necessary because participants did not use a shared dictionary to ensure the same terminology, and so there were multiple cases where the same concern was expressed using different terms (see section~\ref{sec:lexicalOverlapping}). We considered using a dictionary a threat to validity, because it would have influenced the participants and the terms in the dictionary would have suggested them the possible concerns.

The second step of the analysis was the semi-automatic identification of concern occurrences overlapping (formally defined in section~\ref{sec:occurenceSharing}). For each annotated project, we created an XML snapshot which contained a list of all annotation types with the names of program elements annotated by them. These snapshots were created using the Monitor feature of the SSCE NetBeans plugin\footnote{\url{http://github.com/MilanNosal/sieve-source-code-editor}} which inspects source code using standard Java APIs (application programming interfaces). The document was processed by a short Java program which created a report that contained a list of all concerns, for each concern a list of all distinct program elements associated with the given concern, and for each program element a list of participants who annotated the given program element. Finally, this report was analyzed using XPath to obtain the results presented in the following sections.

\subsection{Results}

The number of concerns (annotation types) created by individual participants ranged from 11 to 24, and the number of annotation occurrences from 56 to 140 -- see Table~\ref{tab:summary}.

\begin{table}
\caption{The number of recognized concerns and annotation occurrences per individual subjects} \label{tab:summary}
\centering
\footnotesize
\begin{tabular}{|c|c|c|} \hline
\textbf{Subject} & \textbf{Concerns} & \textbf{Occurrences} \\ \hline \hline
A & 11 & 70 \\ \hline
B & 12 & 56 \\ \hline
C & 24 & 108 \\ \hline
D & 20 & 79 \\ \hline
E & 12 & 56 \\ \hline
F & 14 & 89 \\ \hline
G & 17 & 140 \\ \hline \hline
\textbf{Total (distinct)} & \textbf{46} & \textbf{464} \\ \hline
\end{tabular}
\end{table}

\subsubsection{Concern Sharing}

We constructed a set of distinct concerns $C$, i.e., a union of all sets of the concerns recognized by the participants. The size of this set, i.e., the number of distinct concerns, is 46 (the Total row in Table~\ref{tab:summary}).

More than a half of the concerns (26) was shared by at least two participants. We will call them \textit{shared concerns}. A list of all shared concerns is in Table~\ref{tab:concerns}.

\begin{table*}
\caption{A list of all shared concerns} \label{tab:concerns}
\centering
\footnotesize
\begin{tabular}{|r|l|c|c|c|c|c|} \hline
~ & \textbf{Concern} & \parbox[c][25pt]{60pt}{\centering \textbf{Shared by \textit{n} participants}} & {\centering $EA(c)$} & {\centering $wEA(c)$} & $d_L(K_C)$ & $stdev(K_C)$ \\ \hline \hline
1 & Searching & 6 & 61\% & 84\% & 2.22 & 3.58 \\ \hline
2 & Note editing & 5 & 60\% & 77\% & 4.48 & 3.24 \\ \hline
3 & Note change observing & 5 & 50\% & 67\% & 9.36 & 5.2 \\ \hline
4 & Note presenting & 5 & 21\% & 42\% & 6.67 & 4.14 \\ \hline
5 & Unused code & 4 & 67\% & 85\% & 0.0 & 0.0 \\ \hline
6 & Tagging & 4 & 64\% & 81\% & 4.63 & 3.97 \\ \hline
7 & Persistence & 4 & 41\% & 60\% & 11.2 & 6.18 \\ \hline
8 & Links & 4 & 39\%  & 58\% & 0.5 & 0.5 \\ \hline
9 & Note adding & 4 & 38\% & 63\% & 6.13 & 3.89 \\ \hline
10 & Data model & 4 & 36\% & 63\% & 3.0 & 1.8  \\ \hline
11 & Loading notes & 4 & 35\% & 62\% & 7.75 & 4.76 \\ \hline
12 & Saving notes & 4 & 31\%  & 54\% & 7.39 & 4.23 \\ \hline
13 & GUI & 4 & 26\%  & 41\% & 2.0 & 2.12 \\ \hline
14 & Note deleting & 4 & 18\% & 41\% & 6.25 & 3.83 \\ \hline
15 & UI-model mapping & 4 & 4\% & 7\% & 8.63 & 5.24 \\ \hline
16 & Filter implementation & 3 & 18\% & 36\% & 0.0 & 0.0 \\ \hline
17 & TODO & 3 & 8\% & 15\% & 3.56 & 3.98 \\ \hline
18 & Exceptions & 2 & 100\% & 100\% & 7.0 & 7.0 \\ \hline
19 & Utilities & 2 & 50\% & 67\% & 0.0 & 0.0 \\ \hline
20 & Model change watching & 2 & 17\% & 29\% & 8.0 & 8.0 \\ \hline
21 & Filters management & 2 & 12\% & 21\% & 4.0 & 4.0 \\ \hline
22 & Notes manipulation & 2 & 8\% & 15\% & 4.44 & 4.97 \\ \hline
23 & Questions about code & 2 & 0\% & 0\% & 5.56 & 3.95 \\ \hline
24 & Coding by convention & 2 & 0\% & 0\% & 3.5 & 3.5 \\ \hline
25 & BibTeX & 2 & 0\% & 0\% & 2.0 & 2.0 \\ \hline
26 & Domain entity & 2 & 0\% & 0\% & 4.0 & 4.0 \\ \hline
\end{tabular}
\end{table*}

\subsubsection{Occurrence Sharing}
\label{sec:occurenceSharing}

We define an \textit{annotation occurrence} as a relation $R_{AO}$ between a set of all concerns $C$ and program elements $P$. If and only if a concern annotation for the concern $c\in{C}$ annotates the program element $p\in{P}$, then $(c,p) \in R_{AO}$. $R_{AO}$ relates the programmer's concerns to source code elements, where this relation is expressed using concern annotations.

We define a magnitude $m(c,p)$ of the relation $(c,p) \in R_{AO}$ as a number of different developers who associated the concern $c$ with the given program element $p$. The magnitude $m(c,p)$ is an integer bounded by the interval $[1, n]$, where $n$ is the number of developers who recognized the concern. This means at least one developer has to associate $c$ with $p$ for $(c,p)$ to be included in $R_{AO}$.

A relation of \textit{shared annotation occurrences} $R_{SAO}$ is a subset of $R_{AO}$ such that the ordered pair $(c,p) \in R_{AO}$ belongs also to $R_{SAO}$ if and only if $m(c,p)\geq{2}$, i.e., there are at least two different developers who associated the concern $c$ with the given program element $p$.

We constructed a set of all distinct annotation occurrences with a total size of 464 (as noted in Table~\ref{tab:summary}). 128 of them were shared annotation occurrences -- they were shared by at least two participants.

We define \textit{effective agreement} $EA$ as the number of shared annotation occurrences divided by the total number of annotation occurrences. Now the computation of effective agreement $EA(c)$ for a particular concern $c$ can be described by the following formula:

\begin{equation}
    \label{eq:eacformula}
    EA(c) = \frac{\sum_{p\in{P}}{[(c,p)\in{R_{SAO}}]}}{\sum_{p\in{P}}{[(c,p)\in{R_{AO}}]}} \cdot 100\%
\end{equation}

The formula uses the Iverson bracket notation \cite{Iverson}: $[P]$ is defined to be 1 if $P$ is true, and 0 if it is false.

Overall effective agreement for the whole project can be computed using the following formula:

\begin{equation}
    \label{eq:eaformula}
    EA = \frac{\sum_{c\in{C}}\sum_{p\in{P}}{[(c,p)\in{R_{SAO}}]}}{\sum_{c\in{C}}\sum_{p\in{P}}{[(c,p)\in{R_{AO}}]}} \cdot 100\%
\end{equation}

For our EasyNotes study, the overall effective agreement was 27.59\%. It is possible to see the values of effective agreement for individual concerns in Table~\ref{tab:concerns}. We can consider the effective agreement of a specific concern its quality indicator -- to what extent multiple developers agree on the mapping of this concern to source code elements.

In this extended version of \cite{originalPaper}, we introduce \emph{weighted effective agreement} $wEA(c)$ that puts more weight on annotation occurrences with a higher magnitude. We propose using the following formula to compute $wEA(c)$ for a particular concern:

\begin{equation}
    \label{eq:weacformula}
    wEA(c) = \frac{\sum_{p\in{P}}{([(c,p)\in{R_{SAO}}]\cdot{m(c,p)})}}{\sum_{p\in{P}}{([(c,p)\in{R_{AO}}]\cdot{m(c,p)})}}\cdot{100\%}
\end{equation}
  
We expect the weighted effective agreement to be a better quality indicator for concerns recorded in the code. It prefers annotation occurrences with a higher magnitude, which are more likely to be reused by other developers besides the authors recording the concerns. In Table~\ref{tab:concerns} there are $wEA$ values computed for shared concerns.

In Table~\ref{tab:concerns}, the concerns are ordered by $n$, then by $EA$. Note that if we ordered them by $wEA$ instead of $EA$, some concerns would change their positions. For example, concern no. 11 (Loading notes) would be moved above concern no. 7 (Persistence). This is due to the fact that annotation occurrences of concern no. 11 have a higher average magnitude, signifying they were agreed upon by more developers than in the case of concern no. 7. The $wEA$ should be therefore superior to $EA$.

The $wEA$ (or $EA$) metric can be used to evaluate the overlapping of concerns in source code which are identified by multiple developers. We expect that multiple developers will record concerns in code -- author(s) of the source code, but also its maintainers who went through the process of its comprehension. In such a case, the $wEA$ metric and a threshold value can be used to determine which concerns have the potential for comprehension improvement, and which are likely to be useless for other developers. However, there is a need to experimentally determine the value for this threshold. This belongs to the set of future directions of this research.

\subsubsection{Lexical Overlapping}
\label{sec:lexicalOverlapping}

We also calculated an average Levenshtein edit distance between the set of keywords that were used for each particular shared concern. A Levenshtein distance \cite{levenshtein} of two strings is a number of edit operations (add a character, substitute a character, delete a character) needed to change one of them to the other. Two strings are identical if their Levenshtein distance is 0.

The average Levenshtein distance quantifies the closeness of terms used to express the same concern. Considering $K_C$ to be the set of all keywords for a given concern, we computed its average Levenshtein distance $d_L(K_C)$ and its standard deviation $stdev(K_C)$ using the average Hamming distance formula from \cite{Shutao} adapted for the Levenshtein distance. The Levenshtein distance was computed using the algorithm presented in \cite{levenshteinAlg}.

The computed values are presented in Table~\ref{tab:concerns}. They indicate that even though the concerns were shared, the programmers identified them by various words. Although in some cases (e.g., 'Unused code', or 'Filter implementation') the programmers used the same lexical terms, the overall average Levenshtein distance was 4.7 edits, with the highest values up to 11.2 for the Persistence concern.

The Persistence concern is the boldest example of the nomenclature problem in concern identification. It has the highest Levenshtein distance -- it was recognized by four developers who used the following keywords: 'Persistence', 'NoteIO', 'FileManipulation', and finally the developer G who used two terms for the concern to provide finer granularity -- 'NotesPersistenceFormat' and 'WorkingWithFiles'. Although in principle the developers think about the program using the same concerns, they do not always think in the same terms. This variability indicates that a concern vocabulary could be beneficial for the purposes of recording concerns in team projects. These results indicate a specific case of the vocabulary problem in human-computer communication \cite{furnas}.

\subsubsection{Agreement between Participants}

\begin{figure}
\centering
\includegraphics[width=0.4\hsize]{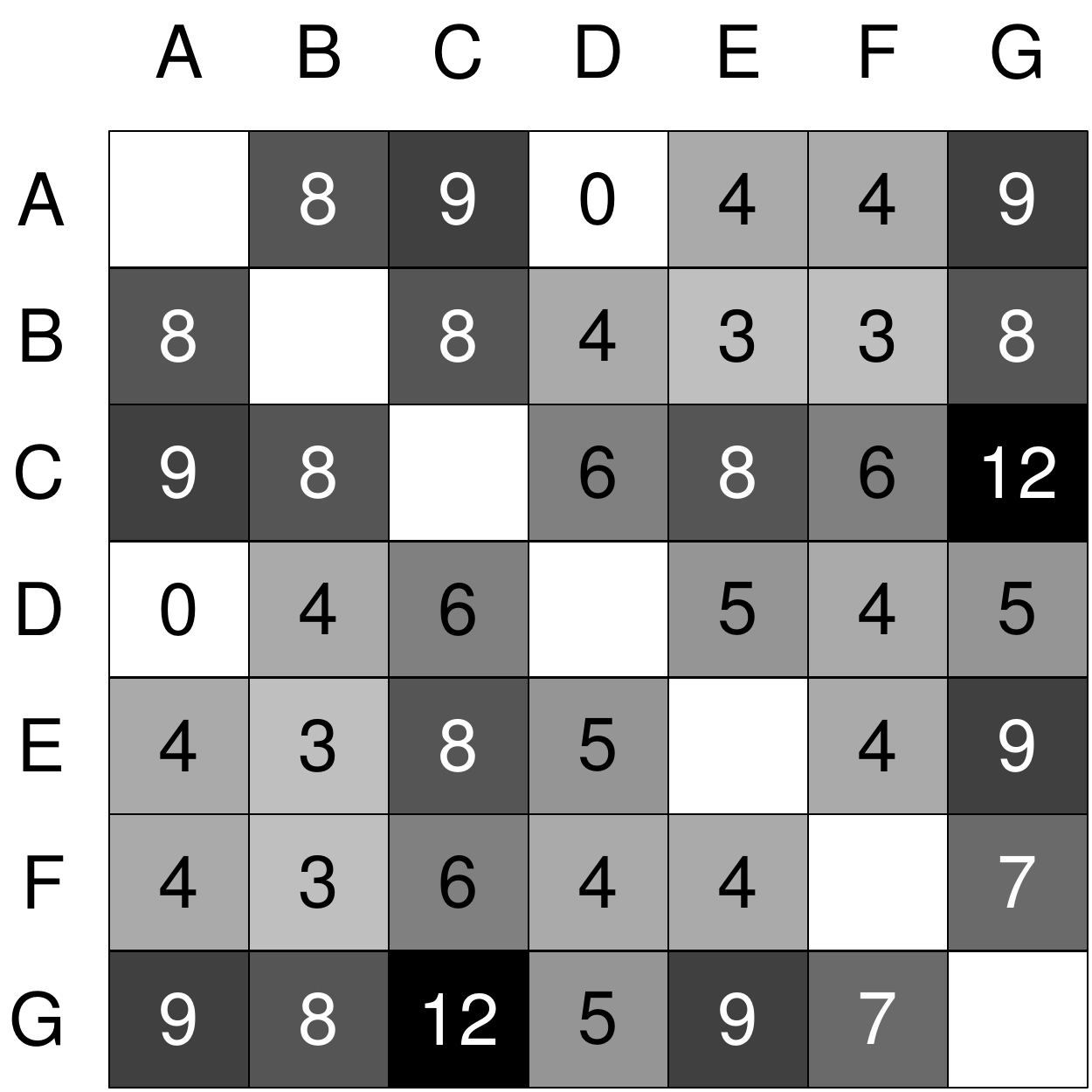}
\caption{Numbers of shared concerns between individual subjects} \label{fig:subjects}
\end{figure}

Fig.~\ref{fig:subjects} shows the number of shared concerns between each pair of participants. We can obtain interesting insights from this matrix. The subjects A and D did not create any common annotation type in our study -- this could be an indication of a huge difference in the mental models of these two people. In fact, D is a former or current supervisor of all other participants except A. On the other hand, G (the author of EasyNotes) shares the most concerns with everyone else. This could mean that the source code author is the best annotator.

A similar matrix was constructed for concern occurrences. Qualitatively, it resembled the annotation type matrix.

\subsubsection{Concern Kinds from the Viewpoint of Shared Concerns}
\label{sec:concernKindsDist}

In section~\ref{sec:kindsConcern}, we specified three kinds of concerns (or concern annotations) -- domain, design, and maintenance. A categorization of the concerns recognized by the participants is in Table~\ref{tab:concernKinds}. Many of the concerns do not have the characteristics of only a single kind. For example, the ``Note change observing'' concern speaks in terms of the problem domain (the term 'Note'), but is mainly used to record the observer pattern. We always selected a concern kind which we considered dominant for the concern.

\begin{table*}
\caption{Kinds of recognized concerns} \label{tab:concernKinds}
\centering
\footnotesize
\begin{tabular}{|l|c||l|c|} \hline
\textbf{Concern} & {\centering \textbf{Kind}} & \textbf{Concern} & {\centering \textbf{Kind}}\\ \hline \hline
Searching & Domain                              & Coding by convention & Design           \\ \hline
Note editing & Domain                           & BibTeX & Domain                         \\ \hline
Note change observing & Design                  & Domain entity & Design                  \\ \hline
Note presenting & Domain                        & Debug logs & Maintenance                \\ \hline
Unused code & Maintenance                       & Currently selected note & Domain        \\ \hline
Tagging & Domain                                & Swing table model & Design              \\ \hline
Persistence & Design                            & Closing notes & Domain                  \\ \hline
Links & Domain                                  & Exit code & Design               \\ \hline                
Note adding & Domain                            & Cancel changes & Design               \\ \hline  
Data model & Design                             & Alternative naming & Design  \\ \hline
Loading notes & Domain                          & Optimization & Design  \\ \hline
Saving notes & Domain                           & Design pattern & Design \\ \hline
GUI & Design                                    & Factory pattern & Design  \\ \hline
Note deleting & Domain                          & Abstract keyword & Maintenance \\ \hline
UI-model mapping & Design                       & Input for searching & Domain \\ \hline
Filter implementation & Design                  & Domain method & Design \\ \hline
TODO & Maintenance                              & Aggregator & Design \\ \hline
Exceptions & Design                             & Handler pattern & Design  \\ \hline
Utilities & Design                              & Template method & Design  \\ \hline
Model change watching & Design                  & Links UI & Design \\ \hline
Filters management & Design                     & Comments & Maintenance \\ \hline
Notes manipulation & Domain                     & Notes sorting & Domain \\ \hline
Questions about code & Maintenance              & Dynamic content panels & Design \\ \hline
\end{tabular}
\end{table*}

In Table~\ref{tab:sharingKinds}, there is a summary of sharing across concern kinds. As we can see, 15 concerns are domain, 25 design and 6 maintenance. This corresponds to 33\%, 54\%, and 13\% of 46 concerns in total. This indicates that developers tend to record their perceptions of design more than of the problem domain and maintenance. We hypothesize that this is because source code identifiers usually already express the domain concern of the program element: e.g., a GUI (graphical user interface) method \texttt{showNote(Note note)}.

\begin{table}
\caption{Sharing of concerns across concern kinds} \label{tab:sharingKinds}
\centering
\footnotesize
\begin{tabular}{|l|c|c|c|} \hline
\textbf{Concern kind} & \textbf{Shared} & \textbf{Not shared} & \textbf{Total} \\ \hline \hline
Domain & 11 (73\%) & 4 (27\%) & 15  \\ \hline
Design & 12 (48\%) & 13 (52\%) & 25  \\ \hline
Maintenance & 3 (50\%) & 3 (50\%) & 6  \\ \hline
\end{tabular}
\end{table}

When we consider a ratio of shared concerns to unshared concerns of a particular kind (Table~\ref{tab:sharingKinds}), we can see that the most shared concern kind are problem domain concerns (73\% of domain concerns were shared). This might indicate that developers are more likely to correctly recognize and identify problem domain concerns in the code. When it comes to code design, it seems that developers tend to look at the code from different perspectives.

\subsubsection{Concern Kinds from the Viewpoint of Developers}
\label{sec:concernKindsDistDevs}

We have calculated a number of particular concern kinds for each developer in particular. The distribution is presented in Table~\ref{tab:developersKinds}. 

\begin{table}
\caption{Distribution of concern kinds according to developers} \label{tab:developersKinds}
\centering
\footnotesize
\begin{tabular}{|c|r|r|r|} \hline
\textbf{Developer} & \textbf{Domain} & \textbf{Design} & \textbf{Maintenance} \\ \hline \hline
A & 7 (64\%) & 2 (18\%) & 2 (18\%)  \\ \hline
B & 7 (58\%) & 4 (33\%) & 1 (8\%)  \\ \hline
C & 14 (58\%) & 7 (29\%) & 3 (13\%)  \\ \hline
D & 5 (25\%) & 11 (55\%) & 4 (20\%)  \\ \hline
E & 6 (50\%) & 6 (50\%) & 0 (0\%)  \\ \hline
F & 1 (7\%) & 11 (79\%) & 2 (14\%)  \\ \hline
G & 8 (47\%) & 8 (47\%) & 1 (6\%)  \\ \hline
\end{tabular}
\end{table}

Multiple participants slightly prefer domain concerns. However, the concerns recognized by developers D (an associate professor) and F (a PhD student) are mostly oriented towards design and maintenance. We have not stated any hypothesis about the distribution of concern kinds recognized by a developer, but examining the possible causes of a particular distribution in the case of a particular developer seems like an interesting future research direction.

\subsection{Threats to Validity}

While some hypotheses were outlined in this section, because this is an exploratory study \cite{Runeson09guidelines}, there is a need to properly quantify and statistically confirm or reject them. This suggests an interesting direction in future research.

In this study, only classes, member variables and methods were annotated. This means that a set of possible program elements (section \ref{sec:occurenceSharing}) was limited to elements of these types.

The participants were unable to mark only a part of a method (a set of statements). We therefore studied overlapping with method granularity. With statement granularity, the results could be different. For instance, if two participants annotated disjoint sets of statements inside one method, the overlap for this particular method would be 0\% instead of 100\%.

The majority of participants had only minor experience with EasyNotes. However, they were given a chance to become familiar with the source code -- both before the study and during it.

We performed the study only on a small-scale Java project. A more extensive study should be conducted in the future.

All participants were previous or current Computer Science students. Furthermore, they were all from the same department. This could make their mental models similar to each other.

\subsection{Conclusion}

We studied mental model overlapping, where parts of the mental model were represented by source code annotations. In our study, about 57\% of all concerns and 28\% of concern occurrences were shared by at least two participants. This means there is potential in recording and subsequent reuse of these data.

Analysis of the average Levenshtein distance between the terms used to identify concerns also showed that a concern dictionary should be used to support comprehension.

An interesting observation is that although in general more design concerns were identified, the most effective sharing was achieved on domain concerns.


\section{Concern Annotations Use Cases}
\label{sec:concernUseCases}

The goal of the second study is to find out how third-party developers, i.e., not the annotation authors, use the concern annotations in the source code if they are available.

\subsection{Method}

We conducted an observational study.

\subsubsection{Materials}
We copied all annotation types shared by at least two subjects (from the first study, see Table~\ref{tab:concerns} for a list) into the EasyNotes project. Thus, all concerns that were recognized by at least two participants were included. Then, for each of these annotation types, we merged all its occurrences (recognized by at least one developer) into the project. The resulting annotations were manually edited for consistency by the EasyNotes author. The resulting source code is published as the commit {\small\texttt{f52872b}}.

\subsubsection{Participants}
There were three participants:
\begin{description}
\item[K] a first-year Computer Science PhD student,
\item[L] a masters degree student with minor industrial experience,
\item[M] a professional developer with 2 years of industrial experience.
\end{description}
None of them had any previous knowledge of EasyNotes.

\subsubsection{Procedure}
First, all annotation types (concerns) were briefly introduced by the EasyNotes author. The subjects were reminded about common feature location possibilities of the NetBeans IDE.

Each participant was given the same task: To add a ``note rating'' (one to five stars) feature to EasyNotes. The fulfillment of this task required a modification of multiple application layers -- from the model to the user interface.

We used a think-aloud method \cite{Runeson09guidelines}, i.e., the participants were kindly requested to comment their thoughts when comprehending the code.

\subsection{Results}

We will now look at typical use cases of concern annotations during our study.

\subsubsection{Confirming Hypotheses}

The most common use of concern annotations was to confirm hypotheses about the code. For example, participant K used the {\small\texttt{\makeatletter{}@\makeatother{}NotesSaving}} annotation to confirm that a particular piece of stream-writing code actually saves notes.

\subsubsection{Feature Location}

In contrast to traditional comments, it is possible to use the Find Usages feature on an annotation type to find all concern occurrences. Our participants were searching for occurrences of the ``filtering'', ``note adding'', ``note saving'' concerns and others. This was considered helpful especially to find whether they did not forget to implement necessary methods for a particular aspect of the note rating feature.

\subsubsection{Non-Obvious Concerns}

The developers also used annotations to obtain new knowledge about the source code. For instance, the UI (user interface) code contained a method used both when adding a new note and when editing an existing one. However, just a note editing concern was obvious from a brief source code inspection. Only thanks to the concern annotation {\small\texttt{\makeatletter{}@\makeatother{}NoteAdding}}, participant M noticed that the code is used for note adding too.

\subsubsection{Elements Relationship}

The subjects noticed that if two or more elements are marked with the same annotation type, there is an implicit relationship between them. For instance, when using the MVC (Model-View-Controller) design pattern, the code in the model marked with a specific annotation is linked with the UI code with the same annotation.

\subsubsection{Maintenance Notes}

The annotation {\small\texttt{\makeatletter{}@\makeatother{}Unused}} marks the methods not used in the rest of the code. This helped the participants to skip them when scanning the code and thus save time.

\subsection{Conclusion}

Concern annotations have several advantages. The participants stated that compared to traditional natural language comments, annotations are much shorter and thus easier to spot. Therefore, they confirmed our expectations discussed in section~\ref{sec:concernAnnotProp}. They are also better structured and usually less ambiguous\footnote{This claim refers to concern annotations in the EasyNotes project, commit {\small\texttt{f52872b}}. In general, an annotation with a single string parameter containing the whole JavaDoc comment could also be considered a concern annotation, but would not be any more structured than the conventional comment. To generalize this advantage, we need to design guidelines or rules on how to define concern annotations (e.g., encourage using enumeration parameters, etc.). We leave this task for our future work.}. The ability to find all usages of a particular concern through standard features present in contemporary IDEs was also appreciated.

Regarding disadvantages, the participant with industrial experience (M) remarked there is a possible scaling problem. Even in a small project like EasyNotes, 26 shared concerns were identified. In large projects, where this number is expected to grow, some sort of concern categorization would definitely be needed. As Java annotations do not support inheritance, marking them with meta-annotations or sorting them to packages are possible solutions.


\section{The Effect of Annotations on Program Maintenance}
\label{sec:effectAnnotations}

We performed a controlled experiment to study the effect of the annotated source code on program comprehension and maintenance.

The guidelines to perform software engineering experiments on human subjects \cite{Ko15practical} were used. To present our findings, the experiment reporting guidelines \cite{Jedlitschka05reporting} were followed. We customized them to the specific needs of this experiment.

\subsection{Hypothesis}

Similarly to Bari\v{s}i\'c et al. \cite{Barisic11quality}, we were interested in correctness, time and confidence.

We hypothesize that the presence of concern annotations in the source code improves program comprehension and maintenance correctness, time and confidence. Thus, we formulate the null and alternative hypotheses:

\textbf{H1$_{\text{null}}$}: The correctness of the results of program comprehension and maintenance tasks on an annotated project = the correctness on the same project without concern annotations.

\textbf{H1$_{\text{alt}}$}: The correctness of the results of program comprehension and maintenance tasks on an annotated project $>$ the correctness on the same project without concern annotations.

\textbf{H2$_{\text{null}}$}: The time to complete program comprehension and maintenance tasks on an annotated project = the time to complete them on the same project without concern annotations.

\textbf{H2$_{\text{alt}}$}: The time to complete program comprehension and maintenance tasks on an annotated project $<$ the time to complete them the same project without concern annotations.

\textbf{H3$_{\text{null}}$}: Participants' confidence of their answers to program comprehension questions on an annotated project = their confidence on the same project without concern annotations.

\textbf{H3$_{\text{alt}}$}: Participants' confidence of their answers to program comprehension questions on an annotated project $>$ their confidence on the same project without concern annotations.

We will statistically test the hypotheses with a confidence level of 95\% ($\alpha=5\%$).

\subsection{Variables}

Now we will define independent variables, i.e., the factors we control, and dependent variables -- the outcomes we measure.

\subsubsection{Independent Variables}

There is only one independent variable -- the presence of concern annotations in the project. The possible values are: yes (``annotated'') and no (``unannotated'').

\subsubsection{Dependent Variables}

The first dependent variable, correctness, was measured as a number of correct answers (or correctly performed tasks) divided by the total number of tasks (5). The tasks are not weighted, each of them is worth one point. The assessment is subjective -- by a researcher.

The second dependent variable is the time to finish the tasks. We are interested mainly in the total time, i.e., a sum of times for all tasks.

Instead of measuring only the time, it is possible to define efficiency as a number of correct tasks and questions divided by time. On one hand, efficiency depends on correctness, which already is a dependent variable. On the other hand, efficiency can deal with participants who fill the answers randomly to quickly finish \cite{Kosar12program}. We decided to use efficiency only as an auxiliary metric to make sure that time differences are still significant even if the correctness is considered.

For each comprehension question, we also asked a subject how confident he/she was on a 3-point Likert scale\footnote{A Likert scale \cite{Likert32technique} is a technique to measure agreement or disagreement with statements. For example, a 3-point Likert scale about frequency looks like: ``always'', ``sometimes'', ``never''. Strictly speaking, a true Likert scale must encompass a set of question-scale pairs. However, we will use a common terminology which calls each such scale ``Likert''.}: from Not at all (1) to Absolutely (3). Since we asked a subject about the confidence equally for each task, we consider it meaningful to calculate the mean confidence, which is the third dependent variable.

\subsection{Experiment Design}

\subsubsection{Materials}

Again, the EasyNotes project was used. This time, we prepared two different versions: with shared concern annotations (as in section \ref{sec:concernUseCases}), and without annotations.

As the project was only scarcely commented, we deleted all traditional source code comments from both versions to remove a potential confounding factor. Only comments for the annotation types themselves were left intact, as we regard them as their integral part. We do not compare concern annotations to traditional comments, since as we noted in section~\ref{sec:concernAnnotations}, concern annotations are not meant to substitute comments.

During this experiment, we used the NetBeans IDE.

\subsubsection{Participants}

We used 18 first-year, master's degree Computer Science students as participants. Carver et al. \cite{Carver10checklist} recommend to integrate software engineering experiments performed on students with teaching goals. We decided to execute the experiment as a part of the Modeling and Generation of Software Architectures course, which contained Java annotations in its curricula.

The course was attended by students focused not only on software engineering, but also on other computer science subfields. Inclusion criteria were set to select mainly students with a prospective future career as professional programmers. Additionally, as EasyNotes is a Java project, a sufficient Java language knowledge was required.

\subsubsection{Design}

When assigning the subjects to groups, we applied a completely randomized design \cite{Wohlin12experimentation}. This means that each group received only one treatment -- either an annotated or an unannotated program -- and the assignment was random. Each participant drew a numbered piece of paper. Subjects with an odd number were assigned to the ``annotated'' group, participants with an even number to the ``unannotated'' one. Our design was thus balanced, with n=9 per group.

\subsubsection{Instruments}

To both guide the subjects and collect the data, we designed an interactive web form\footnote{\url{http://www.jotformeu.com/sulir/sharing-annotations}}. All fields in the form were mandatory, so the participants could not skip any task.

The SSCE NetBeans plugin was used to collect data like session times to an XML file. The participants uploaded this file to the web form at the end of the experiment.

\subsection{Procedure}

The experiment was performed during three lessons in the same week -- the first time with four students, then with 9 and finally with the remaining 5 students. Each session lasted approximately 1.5 hours.

\subsubsection{Training}

At the beginning of the experiment, the users were given 5 minutes to familiarize themselves with EasyNotes from an end-user perspective, without looking at the source code. Then, the session monitoring plugin was briefly introduced.

A short presentation about concern annotations usage in the NetBeans IDE followed. A researcher presented how to show a list of all available concerns, how to use the Find Usages feature on an annotation type and how to navigate from an annotation occurrence to the annotation type.

Just before each of the two maintenance tasks, the researcher presented the participants a running application with the required feature already implemented. This conveniently supplemented natural language task descriptions.

\subsubsection{Tasks}
The experiment comprised of:
\begin{itemize}
\item one additive maintenance task (we will name it \textit{Filter}),
\item three program comprehension questions (\textit{Test}),
\item one corrective maintenance task (\textit{Cite}),
\end{itemize}
in that order. The tasks were formulated as follows:

\begin{description}
\item[Filter] In a running EasyNotes application, load the sample notes and look at the search feature (the down-right part of the window). Try how searching in the note text works (the option ``Search in'': ``text''). Your task will be to add a filter to the EasyNotes application, which will search in the notes title (the option ``title'').
\item[Cite] In the EasyNotes application, there is a field ``Cite as:'' (the down-right window part). Currently, it displays information in the form: \textit{somePubID} where somePubID is the value of the ``PubID:'' field. Your task is to modify the program so the ``Cite as:'' field will display information in the form: \textit{\textbackslash{}cite\{somePubID\}}.
\end{description}
Both tasks were simple, although the \textit{Filter} task was slightly more complex than the latter. It required the creation of a new class with approximately 15 lines of code, whereas the \textit{Cite} task could be accomplished by modifying just one source code line.

The questions asked in the \textit{Test} about program comprehension were:
\begin{description}
\item[Q1] What does the {\small\texttt{runDirMenuItemActionPerformed}} in the class {\small\texttt{easynotes.swingui .EasyNotesFrame}} do?
\item[Q2] How is the class {\small\texttt{easynotes.model.abstractModel.UTFStringComparator}} used in the EasyNotes project?
\item[Q3] What method/s (and in which class) perform(s) note deleting?
\end{description}

Although the tasks were not equal in the expected completion time, the comprehension part was comparably difficult, as none of them required extensive understanding of the code.

\subsubsection{Debriefing}

We also included a question asking to what extent did the subjects use annotations when comprehending the code. Possible answers ranged from Never to Always on a 5-point scale. Finally, the form also contained a free-form question where the participants could describe how the annotations helped them in their own words.

\subsection{Results}

The measured values and their summary is presented in Table~\ref{tab:results}.

Each specific hypothesis considers one independent variable on a nominal scale with two levels (annotated, unannotated) and one dependent variable (either correctness, time or confidence). For each dependent variable, we displayed the values on a histogram and a normal Q-Q plot. None of the variables looked normally distributed, so we used the Mann-Whitney U test as a statistical test for our hypotheses.

\begin{table*}
\caption{The experiment results for individual subjects} \label{tab:results}
\centering
\footnotesize
\setlength{\tabcolsep}{1.5pt}
\begin{adjustwidth}{-1.9cm}{0cm}
\begin{tabular}{|r|cccccr|rrrr|r|cccr|c|} \hline
\multicolumn{17}{|c|}{\textbf{\small The ``annotated'' group}} \\ \hline

\multirow{2}{*}{\textbf{ID}} & \multicolumn{6}{c|}{\textbf{Correctness [true/false]}} & \multicolumn{4}{c|}{\textbf{Time [min]}} & \multirow{2}{44pt}{\centering\textbf{Efficiency [task/min]}} & \multicolumn{4}{c|}{\textbf{Confidence [1-3]}} & \multirow{2}{53pt}{\centering\textbf{Annotations useful? [1-5]}} \\ \cline{2-11} \cline{13-16}

& \textbf{Filter} & \textbf{Cite} & \textbf{Q1} & \textbf{Q2} & \textbf{Q3} & \textbf{Total} & \textbf{Filter} & \textbf{Cite} & \textbf{Test} & \textbf{Total} & & \textbf{Q1} & \textbf{Q2} & \textbf{Q3} & \textbf{Mean} & ~ \\ \hline
1 & 1 & 1 & 1 & 0 & 1 & 80\% & 12.03 & 3.00 & 13.96 & 28.99 & 0.14 & 1 & 2 & 3 & 2.00 & 1 \\ \hline
3 & 0 & 1 & 0 & 0 & 0 & 20\% & 5.23 & 2.81 & 11.13 & 19.17 & 0.05 & 3 & 2 & 3 & 2.67 & 3 \\ \hline
5 & 1 & 1 & 1 & 1 & 1 & 100\% & 17.43 & 3.93 & 6.71 & 28.07 & 0.18 & 3 & 3 & 3 & 3.00 & 4 \\ \hline
7 & 0 & 1 & 1 & 1 & 1 & 80\% & 7.79 & 1.43 & 11.85 & 21.07 & 0.19 & 3 & 3 & 3 & 3.00 & 4 \\ \hline
9 & 1 & 1 & 1 & 0 & 1 & 80\% & 6.72 & 5.86 & 3.87 & 16.45 & 0.24 & 3 & 3 & 2 & 2.67 & 2 \\ \hline
11 & 1 & 1 & 1 & 0 & 1 & 80\% & 8.41 & 4.58 & 4.32 & 17.31 & 0.23 & 3 & 3 & 3 & 3.00 & 4 \\ \hline
13 & 1 & 1 & 0 & 0 & 1 & 60\% & 20.97 & 3.48 & 8.80 & 33.25 & 0.09 & 3 & 2 & 3 & 2.67 & 3 \\ \hline
15 & 1 & 1 & 1 & 0 & 1 & 80\% & 4.64 & 1.91 & 5.22 & 11.77 & 0.34 & 2 & 2 & 3 & 2.33 & 2 \\ \hline
17 & 1 & 1 & 1 & 0 & 1 & 80\% & 25.08 & 6.38 & 6.99 & 38.45 & 0.10 & 2 & 2 & 3 & 2.33 & 4 \\ \hline
\textbf{Median} & \textbf{1} & \textbf{1} & \textbf{1} & \textbf{0} & \textbf{1} & \textbf{80\%} & \textbf{8.41} & \textbf{3.48} & \textbf{6.99} & \textbf{21.07} & \textbf{0.18} & \textbf{3} & \textbf{2} & \textbf{3} & \textbf{2.67} & \textbf{3} \\ \hline
\textbf{Std.dev.} & \textbf{-} & \textbf{-} & \textbf{-} & \textbf{-} & \textbf{-} & \textbf{22.36\%} & \textbf{7.41} & \textbf{1.67} & \textbf{3.57} & \textbf{8.80} & \textbf{0.09} & \textbf{-} & \textbf{-} & \textbf{-} & \textbf{0.35} & \textbf{-} \\ \hline
\hline \multicolumn{17}{|c|}{\textbf{\small The ``unannotated'' group}} \\ \hline

\multirow{2}{*}{\textbf{ID}} & \multicolumn{6}{c|}{\textbf{Correctness [true/false]}} & \multicolumn{4}{c|}{\textbf{Time [min]}} & \multirow{2}{44pt}{\centering\textbf{Efficiency [task/min]}} & \multicolumn{4}{c|}{\textbf{Confidence [1-3]}} & \multirow{2}{53pt}{\centering\textbf{Annotations useful? [1-5]}} \\ \cline{2-11} \cline{13-16}

& \textbf{Filter} & \textbf{Cite} & \textbf{Q1} & \textbf{Q2} & \textbf{Q3} & \textbf{Total} & \textbf{Filter} & \textbf{Cite} & \textbf{Test} & \textbf{Total} & & \textbf{Q1} & \textbf{Q2} & \textbf{Q3} & \textbf{Mean} & ~ \\ \hline
2 & 1 & 1 & 1 & 0 & 1 & 80\% & 2.84 & 4.72 & 10.66 & 18.22 & 0.22 & 3 & 2 & 3 & 2.67 & NA \\ \hline
4 & 0 & 0 & 1 & 0 & 1 & 40\% & 8.76 & 23.45 & 8.10 & 40.31 & 0.05 & 3 & 2 & 3 & 2.67 & NA \\ \hline
6 & 1 & 1 & 1 & 0 & 0 & 60\% & 18.24 & 5.23 & 5.62 & 29.09 & 0.10 & 3 & 2 & 1 & 2.00 & NA \\ \hline
8 & 1 & 1 & 1 & 0 & 1 & 80\% & 6.47 & 5.59 & 11.23 & 23.29 & 0.17 & 3 & 2 & 3 & 2.67 & NA \\ \hline
10 & 1 & 0 & 1 & 0 & 1 & 60\% & 4.82 & 9.64 & 17.50 & 31.96 & 0.09 & 3 & 2 & 3 & 2.67 & NA \\ \hline
12 & 1 & 0 & 1 & 0 & 1 & 60\% & 11.11 & 2.09 & 11.30 & 24.50 & 0.12 & 2 & 2 & 2 & 2.00 & NA \\ \hline
14 & 0 & 1 & 0 & 0 & 1 & 40\% & 30.73 & 7.19 & 5.50 & 43.42 & 0.05 & 2 & 2 & 3 & 2.33 & NA \\ \hline
16 & 1 & 1 & 1 & 0 & 1 & 80\% & 12.56 & 18.39 & 16.07 & 47.02 & 0.09 & 3 & 2 & 3 & 2.67 & NA \\ \hline
18 & 1 & 1 & 1 & 0 & 1 & 80\% & 25.54 & 9.59 & 12.94 & 48.07 & 0.08 & 3 & 2 & 3 & 2.67 & NA \\ \hline
\textbf{Median} & \textbf{1} & \textbf{1} & \textbf{1} & \textbf{0} & \textbf{1} & \textbf{60\%} & \textbf{11.11} & \textbf{7.19} & \textbf{11.23} & \textbf{31.96} & \textbf{0.09} & \textbf{3} & \textbf{2} & \textbf{3} & \textbf{2.67} & \textbf{NA} \\ \hline
\textbf{Std.dev.} & \textbf{-} & \textbf{-} & \textbf{-} & \textbf{-} & \textbf{-} & \textbf{16.67\%} & \textbf{9.56} & \textbf{6.98} & \textbf{4.18} & \textbf{11.06} & \textbf{0.06} & \textbf{-} & \textbf{-} & \textbf{-} & \textbf{0.30} & \textbf{-} \\ \hline
\end{tabular}
\end{adjustwidth}
\end{table*}

\subsubsection{Correctness} \label{sec:correctness}

\begin{figure}
\centering
\begin{subfigure}{0.325\hsize}
\includegraphics[width=\hsize]{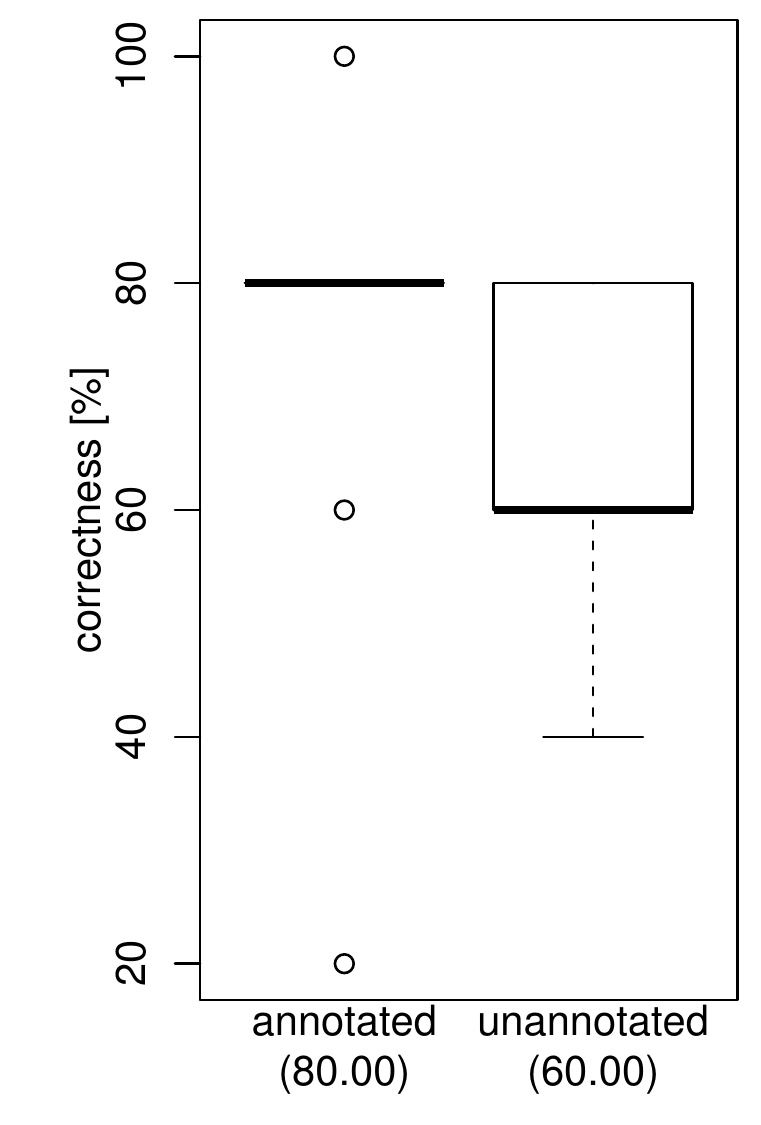}
\caption{correctness} \label{fig:correctness1}
\end{subfigure}
\begin{subfigure}{0.325\hsize}
\includegraphics[width=\hsize]{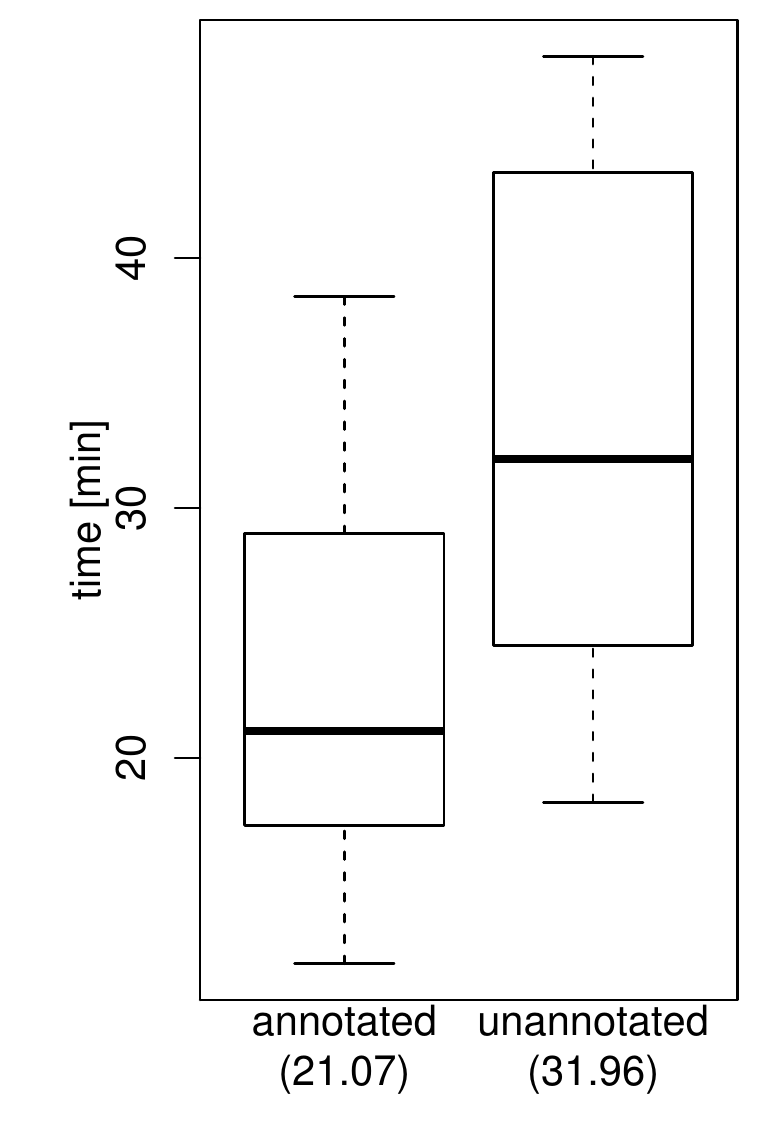}
\caption{time} \label{fig:time1}
\end{subfigure}
\begin{subfigure}{0.325\hsize}
\includegraphics[width=\hsize]{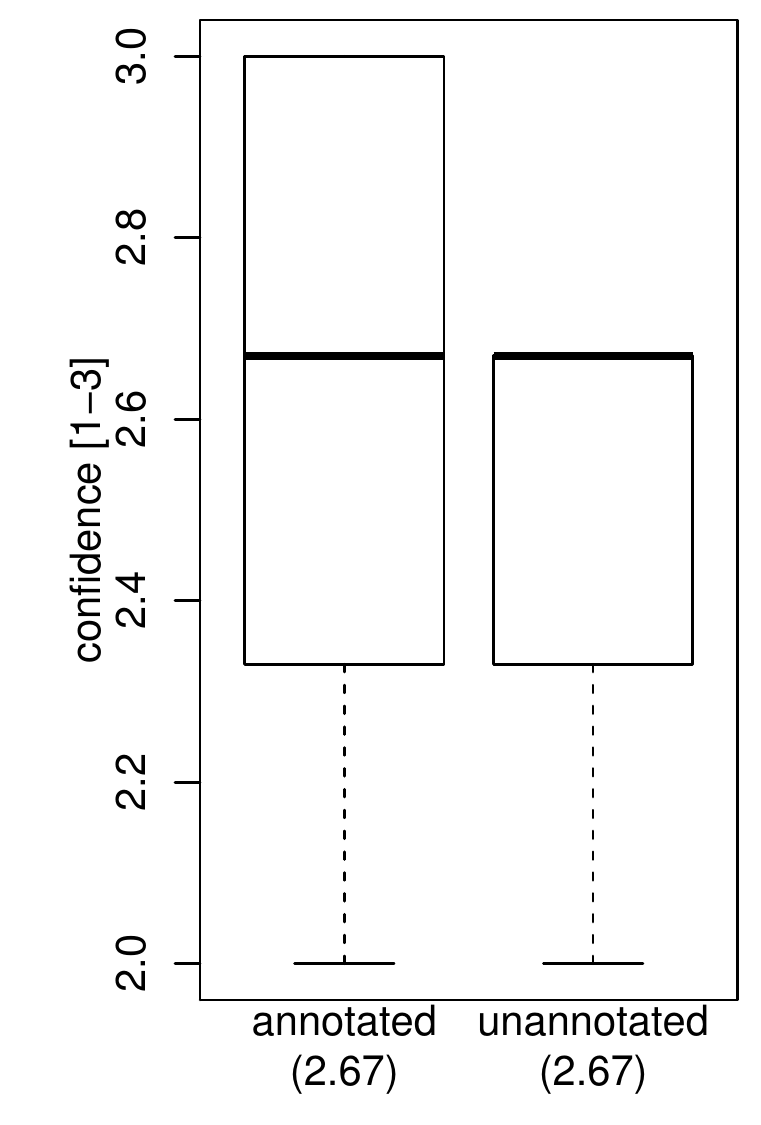}
\caption{confidence} \label{fig:confidence1}
\end{subfigure}
\caption{The results of the controlled experiment.} \label{fig:results1}
\end{figure}

The median correctness for the ``annotated'' was better than for the ``unannotated'' group: 80\% vs. 60\%. See Fig.~\ref{fig:correctness1} for a plot\footnote{The ``correctness'' is slightly different than in the original paper \cite{originalPaper}, as we did not spot less obvious errors in the subjects' code that were identified during the revision of the work for this paper. Because the results for correctness remain insignificant, this does not threaten the validity of the original article.}.

The computed p-value (roughly speaking, the probability that we obtained the data by chance) is 0.0938, which is more than 0.05 (our significance level). This means we accept \textbf{H1$_{\text{null}}$} -- the result is insignificant. We did not prove that the presence of annotations has a positive effect on program comprehension and maintenance correctness.

As we can see in Table~\ref{tab:results} (column Correctness), the most difficult question was Q2. Only two participants answered it correctly -- both from the ``annotated'' group. The class of interest was not used in EasyNotes at all. This fact was noticeable by looking at the {\small\texttt{\makeatletter{}@\makeatother{}Unused}} annotation.

\subsubsection{Time}

The differences in the total time for all tasks between the two groups are graphically depicted in the box plot in Fig.~\ref{fig:time1}. The median time changed from 31.96 minutes for the ``unannotated'' group to 21.07 minutes for the ``annotated'' one, which is a decrease by 34.07\%.

The p-value of time is 0.0252, which is less than 0.05. The difference is statistically significant, therefore we reject \textbf{H2$_{\text{null}}$} and accept \textbf{H2$_{\text{alt}}$}. The presence of concern annotations improves the program comprehension and maintenance time.

It is possible to see from Table~\ref{tab:results} (column Time) that the median time for each individual task was better for the ``annotated'' group. The most prominent difference was for the task \textit{Cite}. This can be due to the fact that the project contained the concern annotation {\small\texttt{\makeatletter{}@\makeatother{}Citing}} which helped the participants find the relevant code quickly.

The median of efficiency, which we defined as the number of correctly performed tasks (and answers) divided by total time, raised by 89.76\% (p=0.0313). This means the time improvement is significant even if we take correctness into account.

\subsubsection{Confidence}

The median of mean confidence is the same for both groups (2.67), as obvious from Table~\ref{tab:results}, column Confidence / Mean and Fig.\ref{fig:confidence1}. The p-value is 0.1710 ($>$0.05) and therefore we accept \textbf{H1$_{\text{null}}$}. The effect of concern annotations on confidence was not demonstrated.

Looking at the individual questions, Q2 was clearly perceived the most difficult. This corresponds with our previous finding regarding correctness. An interesting fact is that no participant in the ``unannotated'' group was confident enough to select level 3 (Absolutely), while in the ``annotated'' group, there were 4 such subjects and two of them answered correctly.

\subsubsection{Other Findings}

As seen from Table~\ref{tab:results}, column ``Annotations useful?'', concern annotations were perceived relatively helpful by the participants (median 3 on a 5-point scale).

Answers to a free-form question asking how specifically the annotations were useful included: faster orientation in a new project, faster searching (mainly through Find Usages on annotations), less scrolling. The participants also stated: ``they helped me to understand some methods'', ``annotations could be perfect, but I would have to get used to them''.

\subsection{Threats to Validity}

To analyze threats to validity of this experiment, we used also \cite{Neto14threats} as guidelines.

\subsubsection{Construct Validity}

Like Kosar et al. \cite{Kosar12program}, we compensated the students with points for the participation in the experiment, which increased their enthusiasm. Unlike them, we did not reward the participants with bonus points for good results because our experiment spanned several days with the same tasks and this would motivate the students to discuss the task details with classmates -- it could negatively affect the construct validity (interaction between subjects). Furthermore, the participants were explicitly told not to share experiment details with anyone. Therefore, we do not consider executing the experiment in three separate sessions an important validity threat.

To measure confidence, we used only a 3-point Likert scale. This decision was not optimal. Because subjects rarely selected the value of 1 (which can be interpreted as guessing an answer), there were only two values left. This could be one of the reasons we did not find a statistically significant difference in confidence.

\subsubsection{Internal Validity}

We divided the subjects into groups randomly. Another possibility was a quasi-experiment (nonrandom assignment), i.e., to divide the subjects evenly according to a co-factor such as their programming experience. However, random assignments tend to have larger effect sizes than quasi-experiments \cite{Kampenes09systematic}.

Full pilot testing with third-party participants was not performed, we only tested the comprehension questions on one of the researchers. We rejected 3 out of the 6 prepared questions because we considered them too difficult and ambiguous. Despite this, all tasks were either completed by almost all of the participants (Filter, Q1, Q3, Cite) or by almost none (Q2) during the experiment. This negatively affected the results for the ``correctness'' variable.

During the actual experiment, we used a native language (Slovak) version of the web form to eliminate the effect of English knowledge on the results. While the source code was in English, this did not present a validity threat since all participants had at least a medium level (3 on a 5-point scale) of English knowledge, as stated by the subjects on the form.

\subsubsection{External Validity}

We invited only students to our experiment, no professional developers. We can consider this fact positively -- concern annotation consumers are expected to be mainly junior developers, whereas potential annotation creators are mostly senior developers. Furthermore, some students may already work in companies during their study.

EasyNotes is a small-scale Java project -- around 3 KLOC (thousands of lines of code), including annotations. The effect of concern annotations on larger programs should be investigated.

\subsubsection{Reliability}

The concern annotations training (tutorial) was presented manually by a researcher. However, there were two independent researchers who took turns.

The experiment is replicable, as we published the data collection form (\texttt{\small\url{http://www.jotformeu.com/sulir/sharing-annotations}}) which contains both the guidelines and links to the materials (two versions of the EasyNotes project).

\subsubsection{Conclusion Validity}

A small number of subjects (n=9 per group) is the most important conclusion validity threat. If we used a paired design (to assign both treatments to every subject), we could easily reach n=18. However, the participants would quickly become familiar with EasyNotes and the second set of tasks would be affected by their knowledge.

\subsection{Conclusion}

We successfully confirmed the hypothesis that concern annotations have a positive effect on program comprehension and maintenance time. The group which had concern annotations available in their project reached the time more than 34\% shorter than the group without them (p $<$ 0.05).

On the other hand, we did not discover a significant change in correctness and confidence. The difference was neither negative (which could mean the annotations are confusing because of their ``visual noise'') nor positive and it was probably a result of the discussed validity threats.


\section{Replication of the Controlled Experiment}
\label{sec:replication}

While replication is a crucial scientific method, less than 18\% of controlled experiments in software engineering are replications \cite{Sjoeberg05survey}. In this extended version of \cite{originalPaper}, we decided to repeat the experiment with slightly modified conditions, i.e., to perform a differentiated replication \cite{Wohlin12experimentation}.

\subsection{Method}

The hypotheses, the independent variable, dependent variables (except confidence), comprehension questions and maintenance tasks were exactly the same as in the original experiment. The recruitment, demographic characteristics of the participants, training and execution details were different. In the following sections, we report the modified aspects.

\subsubsection{Dependent Variables}
As a 3-point Likert scale for confidence was shown to be too narrow in the original experiment, we decided to change it to a 5-point scale. This allows the subjects to more precisely distinguish between the confidence levels.

\subsubsection{Recruitment}

This time, we concentrated on developers with industrial experience. Besides industrial experience, the main requirement was to have at least some Java knowledge (so that the comprehension would not be affected by an unfamiliar programming language).

We contacted the developers whom we knew directly by e-mail, or in person. Additionally, we distributed an announcement using a social network and an Internet forum. The participation was voluntary and no compensation was provided.

\subsubsection{Participants}

20 subjects decided to participate in the experiment. However, one participant completed only the first task, so we excluded him completely. During the analysis of time results for 19 subjects, we found an outlier. After contacting the person, we found out that he was disrupted during the experiment. We decided to exclude this subject, too.

Next, we present the results for 18 subjects who participated in the whole experiment. We divided them into two groups -- ``annotated'' and ``unannotated'' -- based on their IDs. Each group contained 9 participants.

\subsubsection{Demographic Characteristics} \label{s:demographic}

The average age of the subjects was 26.8 years. All participants except for one were men. Their full-time programming experience ranged from 0 (mostly students with part-time jobs) to 15 years, with a mean of 2.7 years.

We also included an optional question about the participant's company in the form. Overall, the subjects were working in at least 6 companies, so the selection is not biased towards any domain or work practices.

\subsubsection{Procedure}

The participants were sent a link to the web form\footnote{\url{http://www.jotformeu.com/sulir/sharing-annotations}}, which contained all necessary information. They could execute the experiment in their natural settings -- at home. Therefore, the training example was not presented by a researcher. Instead, the web form contained a detailed tutorial. A participant was guided to try the IDE features related to the concern annotations himself, according to the examples presented in the tutorial.

After the completion of the experiment, every subject uploaded the zipped project and the XML file with time data to the web form. We then checked the correctness of each implementation.

\subsection{Results}

The results of the experiment replication are in Table~\ref{tab:results2}.

\begin{table*}
\caption{The detailed results of the experiment replication.} \label{tab:results2}
\centering
\footnotesize
\setlength{\tabcolsep}{1.5pt}
\begin{adjustwidth}{-1.9cm}{0cm}
\begin{tabular}{|r|cccccr|rrrr|r|cccr|c|} \hline
\multicolumn{17}{|c|}{\textbf{\small The ``annotated'' group}} \\ \hline

\multirow{2}{*}{\textbf{ID}} & \multicolumn{6}{c|}{\textbf{Correctness [true/false]}} & \multicolumn{4}{c|}{\textbf{Time [min]}} & \multirow{2}{44pt}{\centering\textbf{Efficiency [task/min]}} & \multicolumn{4}{c|}{\textbf{Confidence [1-5]}} & \multirow{2}{53pt}{\centering\textbf{Annotations useful? [1-5]}} \\ \cline{2-11} \cline{13-16}

& \textbf{Filter} & \textbf{Cite} & \textbf{Q1} & \textbf{Q2} & \textbf{Q3} & \textbf{Total} & \textbf{Filter} & \textbf{Cite} & \textbf{Test} & \textbf{Total} & & \textbf{Q1} & \textbf{Q2} & \textbf{Q3} & \textbf{Mean} & ~ \\ \hline
19 & 1 & 1 & 1 & 0 & 1 & 80\% & 11.94 & 5.48 & 35.95 & 53.37 & 0.07 & 5 & 4 & 4 & 4.33 & 2 \\ \hline
21 & 1 & 1 & 1 & 1 & 1 & 100\% & 19.87 & 14.49 & 9.30 & 43.66 & 0.11 & 4 & 5 & 5 & 4.67 & 3 \\ \hline
23 & 1 & 1 & 1 & 0 & 1 & 80\% & 5.97 & 23.20 & 6.69 & 35.86 & 0.11 & 5 & 4 & 5 & 4.67 & 4 \\ \hline
27 & 1 & 1 & 1 & 1 & 1 & 100\% & 10.67 & 5.01 & 28.90 & 44.58 & 0.11 & 4 & 3 & 5 & 4.00 & 3 \\ \hline
29 & 1 & 1 & 1 & 0 & 1 & 80\% & 19.80 & 3.22 & 7.12 & 30.14 & 0.13 & 4 & 4 & 4 & 4.00 & 3 \\ \hline
31 & 1 & 1 & 0 & 0 & 1 & 60\% & 12.17 & 3.80 & 6.08 & 22.05 & 0.14 & 4 & 5 & 5 & 4.67 & 3 \\ \hline
51 & 0 & 1 & 1 & 0 & 1 & 60\% & 9.12 & 3.59 & 6.29 & 19.00 & 0.16 & 4 & 5 & 5 & 4.67 & 4 \\ \hline
53 & 1 & 1 & 1 & 1 & 1 & 100\% & 9.07 & 13.43 & 9.00 & 31.50 & 0.16 & 5 & 5 & 5 & 5.00 & 3 \\ \hline
55 & 1 & 1 & 1 & 0 & 1 & 80\% & 2.83 & 3.82 & 5.75 & 12.40 & 0.32 & 4 & 4 & 4 & 4.00 & 1 \\ \hline
\textbf{Median} & \textbf{1} & \textbf{1} & \textbf{1} & \textbf{0} & \textbf{1} & \textbf{80\%} & \textbf{10.67} & \textbf{5.01} & \textbf{7.12} & \textbf{31.50} & \textbf{0.13} & \textbf{4} & \textbf{4} & \textbf{5} & \textbf{4.67} & \textbf{3} \\ \hline
\textbf{Std.dev.} & \textbf{-} & \textbf{-} & \textbf{-} & \textbf{-} & \textbf{-} & \textbf{15.63\%} & \textbf{5.67} & \textbf{7.01} & \textbf{11.34} & \textbf{13.32} & \textbf{0.07} & \textbf{-} & \textbf{-} & \textbf{-} & \textbf{0.37} & \textbf{-} \\ \hline
\hline \multicolumn{17}{|c|}{\textbf{\small The ``unannotated'' group}} \\ \hline

\multirow{2}{*}{\textbf{ID}} & \multicolumn{6}{c|}{\textbf{Correctness [true/false]}} & \multicolumn{4}{c|}{\textbf{Time [min]}} & \multirow{2}{44pt}{\centering\textbf{Efficiency [task/min]}} & \multicolumn{4}{c|}{\textbf{Confidence [1-5]}} & \multirow{2}{53pt}{\centering\textbf{Annotations useful? [1-5]}} \\ \cline{2-11} \cline{13-16}

& \textbf{Filter} & \textbf{Cite} & \textbf{Q1} & \textbf{Q2} & \textbf{Q3} & \textbf{Total} & \textbf{Filter} & \textbf{Cite} & \textbf{Test} & \textbf{Total} & & \textbf{Q1} & \textbf{Q2} & \textbf{Q3} & \textbf{Mean} & ~ \\ \hline
20 & 1 & 1 & 0 & 0 & 0 & 40\% & 7.88 & 15.09 & 49.71 & 72.68 & 0.03 & 5 & 3 & 4 & 4.00 & 1 \\ \hline
22 & 1 & 1 & 1 & 0 & 1 & 80\% & 4.71 & 3.19 & 11.93 & 19.83 & 0.20 & 5 & 2 & 5 & 4.00 & 1 \\ \hline
24 & 1 & 1 & 1 & 1 & 1 & 100\% & 6.78 & 6.12 & 16.62 & 29.52 & 0.17 & 4 & 3 & 5 & 4.00 & 1 \\ \hline
26 & 1 & 0 & 0 & 0 & 1 & 40\% & 4.68 & 5.61 & 14.23 & 24.52 & 0.08 & 4 & 4 & 5 & 4.33 & 3 \\ \hline
30 & 1 & 0 & 1 & 0 & 0 & 40\% & 2.53 & 9.78 & 5.00 & 17.31 & 0.12 & 5 & 4 & 4 & 4.33 & 1 \\ \hline
32 & 1 & 1 & 1 & 0 & 1 & 80\% & 9.27 & 2.29 & 17.87 & 29.43 & 0.14 & 4 & 5 & 5 & 4.67 & 1 \\ \hline
50 & 1 & 1 & 0 & 0 & 0 & 40\% & 5.37 & 1.71 & 11.58 & 18.66 & 0.11 & 5 & 4 & 3 & 4.00 & 1 \\ \hline
52 & 1 & 1 & 1 & 0 & 0 & 60\% & 7.02 & 8.22 & 17.39 & 32.63 & 0.09 & 5 & 4 & 4 & 4.33 & 1 \\ \hline
54 & 1 & 1 & 0 & 0 & 1 & 60\% & 4.21 & 8.79 & 4.85 & 17.85 & 0.17 & 4 & 5 & 5 & 4.67 & 1 \\ \hline
\textbf{Median} & \textbf{1} & \textbf{1} & \textbf{1} & \textbf{0} & \textbf{1} & \textbf{60\%} & \textbf{5.37} & \textbf{6.12} & \textbf{14.23} & \textbf{24.52} & \textbf{0.12} & \textbf{5} & \textbf{4} & \textbf{5} & \textbf{4.33} & \textbf{1} \\ \hline
\textbf{Std.dev.} & \textbf{-} & \textbf{-} & \textbf{-} & \textbf{-} & \textbf{-} & \textbf{22.36\%} & \textbf{2.08} & \textbf{4.25} & \textbf{13.34} & \textbf{17.30} & \textbf{0.05} & \textbf{-} & \textbf{-} & \textbf{-} & \textbf{0.28} & \textbf{-} \\ \hline
\end{tabular}
\end{adjustwidth}
\end{table*}

\subsubsection{Correctness}

\begin{figure}
\centering
\begin{subfigure}{0.325\hsize}
\includegraphics[width=\hsize]{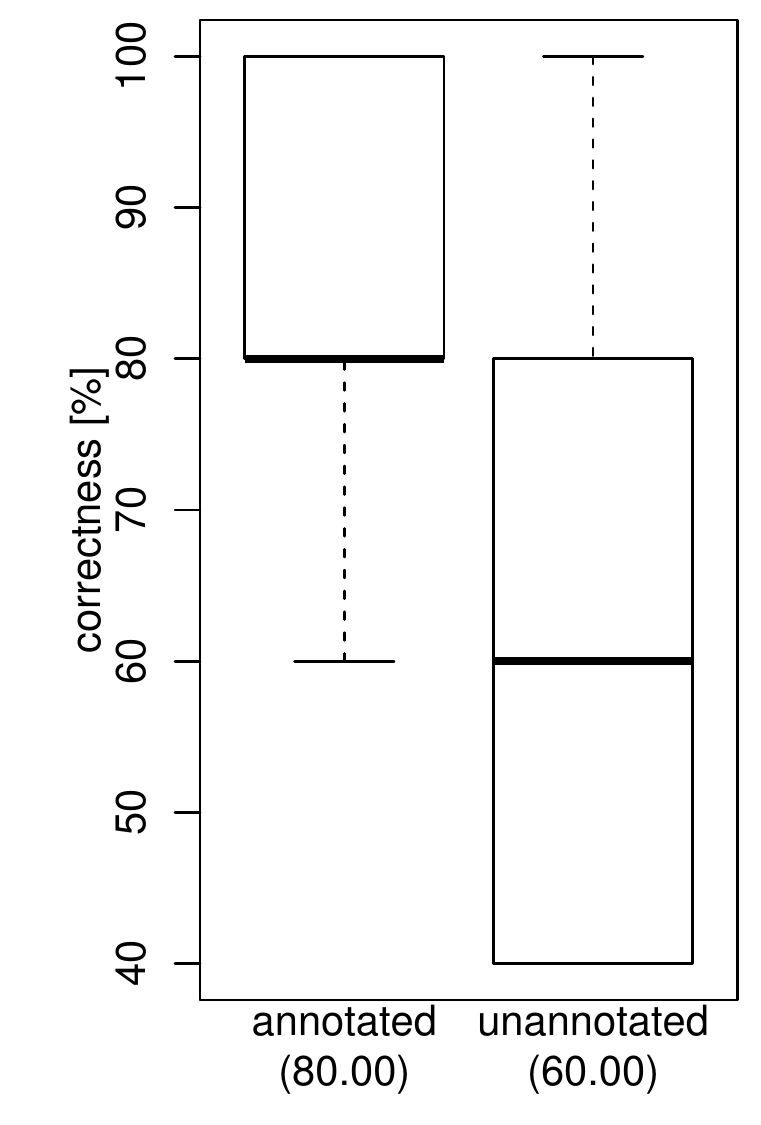}
\caption{correctness} \label{fig:correctness2}
\end{subfigure}
\begin{subfigure}{0.325\hsize}
\includegraphics[width=\hsize]{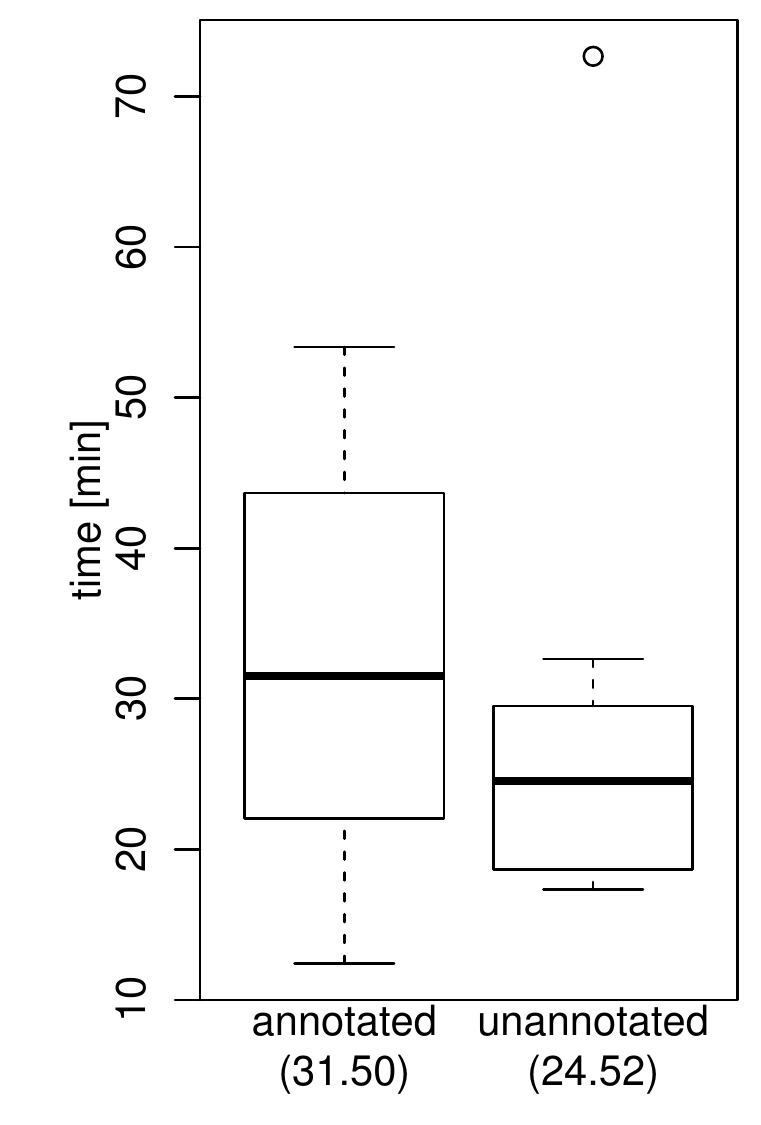}
\caption{time} \label{fig:time2}
\end{subfigure}
\begin{subfigure}{0.325\hsize}
\includegraphics[width=\hsize]{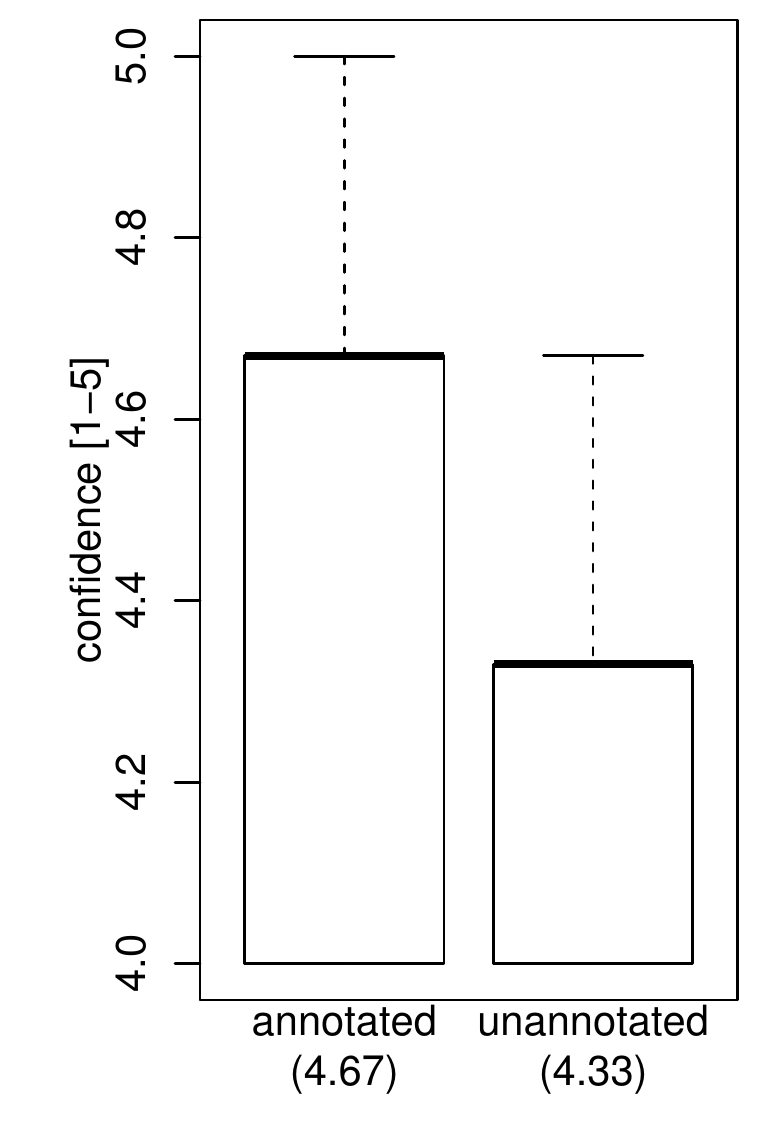}
\caption{confidence} \label{fig:confidence2}
\end{subfigure}
\caption{The results of the experiment replication.} \label{fig:results2}
\end{figure}

The correctness for the ``annotated'' group was better: the median was 80\%, compared to the ``unannotated'' one, which was 60\% (see Fig.~\ref{fig:correctness2}). The p-value is 0.0198, which is less than 0.05. Therefore, the result is statistically significant. In this replication, we reject \textbf{H1$_{\text{null}}$} and accept \textbf{H1$_{\text{alt}}$}.

\subsubsection{Time}
The time for the ``annotated'' group was now about 28\% higher -- see Fig.~\ref{fig:time2}. However, the result is statistically insignificant, since p=0.1487. We therefore cannot claim any difference and accept \textbf{H2$_{\text{null}}$} in the replication.

If we look at the results in terms of efficiency, i.e., the number of correct tasks per minute, the annotated group performed slightly better (Table~\ref{tab:results2}, column Efficiency). Nevertheless, the p-value is 0.3652, so the difference is insignificant.

\subsubsection{Confidence}

The confidence in answering program comprehension questions was almost 8\% better for the ``annotated'' group (see Fig.~\ref{fig:confidence2}). The p-value is 0.1417, so the result is insignificant and we accept \textbf{H3$_{\text{null}}$}.

\subsubsection{Other Findings}
Now we report some of the free-form answers of the subjects:
\begin{itemize}
\item ``It simplified the orientation in code.''
\item ``It took me a while to understand the meaning of the given annotations.''
\item ``Annotations helped me to relatively quickly identify interesting code parts, especially the annotations close to the problem domain (e.g., note deleting).''
\item ``If the architecture was more separated, annotations like Filter and UI would be very useful.''
\item ``Annotations served as a correctness assurance to me.''
\end{itemize}

\subsection{Threats to Validity}

\subsubsection{Internal Validity}

Since the design of the experiment was almost the same as in the original experiment, no new internal validity threats were introduced.

\subsubsection{External Validity}

In the original experiment, only students participated in the experiment. In this replication, all participants had industrial experience.  We therefore raised the external validity considerably. Furthermore, the participants were not from one company, so they represented a variety of domains.

\subsubsection{Reliability}

Instead of relying on a researcher to train the subjects, we provided them a tutorial in a textual form. It was exactly the same for all participants, which has a positive impact on reliability.

\subsection{Conclusion}

Table~\ref{tab:comparison} summarizes the results of both the original experiment and the replication. If we look at statistically significant results, we come to the following conclusions:
\begin{itemize}
\item In the original experiment, the time was improved by about a third.
\item In the replication, the correctness was improved by more than a third.
\end{itemize}

\begin{table}
\caption{A comparison of the original experiment and the replication. The ``change'' means the difference between the annotated and unannotated group. Significant results are highlighted in bold.} \label{tab:comparison}
\centering
\footnotesize
\begin{tabular}{|l||r|l|r|l|r|l|} \hline
~ & \multicolumn{2}{c|}{\textbf{Correctness}} & \multicolumn{2}{c|}{\textbf{Time}} & \multicolumn{2}{c|}{\textbf{Confidence}} \\ \cline{2-7}
~ & change & p-value & change & p-value & change & p-value \\ \hline \hline
\textbf{Original} & +33.3\% & 0.0938 & \textbf{$-$34.1\%} & \textbf{0.0252} & 0\% & 0.1710 \\ \hline
\textbf{Replication} & \textbf{+33.3\%} & \textbf{0.0198} & +28.5\% & 0.1487 & +7.9\% & 0.1417 \\ \hline
\end{tabular}
\end{table}

The correctness was improved also in the first experiment, although the difference was statistically insignificant. The only negative change is the increase of time in the replication. However, it is not statistically significant and it is likely the difference was only by chance. Another possible reason is that users without annotations often submitted incorrect results quickly, while the ``annotated'' group checked their submissions more carefully.

Regarding confidence, no important difference was shown in any of the experiments. We widened the scale from a 3-point in the first experiment to a 5-point in the replication. Although the difference is small and insignificant, the confidence results incline positively to the ``annotated'' group.


\section{Related Work}
\label{sec:related}

There are multiple challenges in source code concern (intent) preservation. For a more broader overview of them we refer the reader to our survey presented in \cite{vranicIntent}. In this section, we will focus on works related to concern annotations with regard to program comprehension improvement.

\subsection{Mental Model Overlapping}

Revelle et al. \cite{Revelle05understanding} also studied concern overlapping. However, their study included only two concern sets, compared to our seven. Their results are positive, too. This further confirms our hypothesis that it is possible to share mental models.

\subsection{Annotations}

Annotations are a program-level declarative marking technique \cite{fedcsisAnnLan}. They are usually used as an implementation technique for pure embedding \cite{languageUntangled}, since they can be considered an alternative to embedded domain-specific language (DSL) implementation techniques \cite{apeg,neverlang,boundedSeas}. Today, another one of the most common applications of annotations is configuration. A programmer marks selected source code elements with appropriate annotations. Later, they are processed either by an annotation processor (during compilation) or by reflection (at runtime). For example, annotations can be used to define concrete syntax in parser generators \cite{Poruban10annotation}, to declare references between elements in a domain-specific language \cite{Lakatos13declarative}, or for memory management \cite{Stancu15safe}. This way, annotations can indirectly modify the annotated program semantics. In contrast, concern annotations utilize annotations just as clues for a programmer which are processed by an IDE when needed. If annotations used by some tool record information about the problem domain of the application (e.g., as is in case of feature model-driven generation \cite{vranicFeature}), they can be beneficial for program comprehension as well.

In Java, annotations can be only applied to packages, classes, member variables, methods and parameters. \makeatletter{}@Java\makeatother{} \cite{Cazzola14java}, an extension to the standard Java language, brings annotations below the method level. It allows the marking of individual source code statements like assignments, method calls, conditions and loops with annotations. This could be useful to capture e.g., algorithmic design decisions with concern annotations.

\subsubsection{Domain Concerns}

Domain annotations (sometimes called semantic annotations \cite{socSemanticAnnotations}) are used mainly in aspect oriented programming to deal with the fragile pointcut problem. Kiczales et al. \cite{socKiczales} recommend using them to express semantic properties of the source code that can later be used to bind some code to annotated program elements by the weaving process. Meffert \cite{designPatternsMeffert} uses semantic annotations to record semantic concerns of the code. He presents a recommendation tool that analyzes source code structure and its concerns to propose design pattern application.

Ji et al. \cite{Ji15maintaining} recommend marking the presence of individual features in source code via specially formatted comments. Using a simulation study, they showed that the cost of adding and maintaining such marks is negligible. This suggests that using our approach, which captures also non-feature concerns and uses Java annotations, could be comparably beneficial.

\subsubsection{Design Annotations}

Design annotations are used to explicitly express design decisions concerning annotated program elements. Hedin \cite{attExtensionHedin} uses attribute extensions (structured comments corresponding with annotations) for design pattern formalization. Each design pattern defines roles for program elements that implement it. Hedin suggests using these attribute extensions to record design pattern instances in the code. Sabo et al. \cite{Sabo09preserving} use design annotations to preserve the design patterns in source code during the development and evolution of the system. They present a tool that can check whether a design pattern was not violated by evolution. Kajsa et al. \cite{designPatternsSupportNavrat} build their work on the work of Sabo et al., but add tool support for design pattern artifacts generation and also for design pattern evolution.

\subsubsection{Maintenance Notes}

Developers often write ``TODO comments'' like {\small\texttt{// TODO: fix this}} to mark parts of source code which need their attention \cite{Storey08todo}. IDEs can then try to parse and display these notes in a task list window. Our approach is more formal, as annotations are a part of the standard Java grammar and can be parsed unambiguously. Furthermore, it is possible to distinguish between multiple maintenance note types through individual annotation types.

TagSEA \cite{Storey06shared} uses a special source code comment format to tag source code elements with maintenance notes and navigation hints. However, their approach relies on a plugin to search and filter the tags.

\subsection{Concerns Recorded in Identifiers}

Identifiers identify program elements using problem domain terms (or possibly in design terms; e.g., index, Observer, etc.). Deissenboeck et al. \cite{conciseNaming} notes that 33\% of all tokens in the source code are identifiers, and they create up to 70\% of the source code characters (measurements of the contemporary version of the Eclipse project). Although identifiers can be used to encode all concerns, annotations provide much better scalability \cite{CepaAESD}.

\subsection{Concern Annotations Utilization}

Our motivation for using annotations for recording concerns is based on the fact that annotations are first-class citizens of the language and there are standard tools for their parsing \cite{fedcsisAnnLan}. This provides a great potential for concern annotations reuse, and for implementing tooling for their utilization. For example, in \cite{Poruban14leveraging}, we used concern annotations as one of the ways to perform source code projections. Projections allow looking at the source code from multiple different perspectives. For example, an IDE plugin can filter all code marked with a specific annotation type and display it in an editable window -- even if it is spread across multiple files.

Niu et al. \cite{Niu11faceted} propose the application of HFC (hierarchical faceted categories) on source code. They enable categorizing source code fragments into hierarchical categories on different facets. Their approach requires specialized tools whereas we use source code annotations which have standard IDE support.


\section{Conclusion and Future Work}
\label{sec:conclusion}

In this paper, we presented an idea of using Java source code annotations to capture and share parts of programmers' mental model, namely the concerns (intents), hence the name, \textit{concern annotations}. Each concern (e.g., ``searching'', ``editing'', ``GUI code'', ``unused code'') is implemented as an annotation type. Subsequently, all classes, methods and fields relevant to that concern are marked with the given annotation. We defined the metrics \textit{effective agreement} and \textit{weighted effective agreement} to measure the concern overlap of multiple developers. Two case studies, one experiment and its replication were conducted to assess the practical implications of this approach.

A precondition to the usefulness of sharing concern annotations among multiple developers is that their mental models at least partially overlap. In the first study, we showed that this precondition holds. More than a half of the concerns created by one of seven developers in our study was recognized by at least two of them. More than 1/4 of concern occurrences (locations in source code where a particular annotation is used) were shared by at least two participants.

In the second study, we discovered that concern annotations are particularly useful to confirm hypotheses about the code, locate the features, find out non-obvious concerns which a method fulfills, and discover hidden relationships between elements. Concern annotations can also be used as a replacement of traditional TODO comments.

The first controlled experiment, which used only student subjects, showed there is a statistically significant improvement of development time when performing program comprehension and maintenance tasks on a small-scale Java project. The group which had an annotated version of the same program available, consumed 1/3 less time than the group which did not have concern annotations present in the source code.

A differentiated replication of the experiment was performed, now focused on professional programmers. A statistically significant improvement of question and task correctness was reached for the ``annotated'' group. While the time was slightly worse, the difference was statistically insignificant.

In our studies, the source code was commented scarcely or not at all. An interesting future comparison would consider an annotated program versus a program without annotations, but in both cases with high-quality traditional source code comments (this set-up would better reflect the real world scenario).

Studying the difference of mental models between individual developers, like the causality between the developer's experience and his preferred annotation types, is another interesting future research direction.

\section*{Acknowledgment}

This work was supported by project KEGA No. 019TUKE-4/2014 Integration of the Basic Theories of Software Engineering into Courses for Informatics Master Study Programmes at Technical Universities -- Proposal and Implementation.

\section*{References}
\bibliography{bibliography}

\listoftodos

\end{document}